%% file: CWI_arxiv.tex
\documentclass[11pt]{article} 

\usepackage{times}
\usepackage{comment}
\usepackage{bm}
\usepackage{natbib}
\usepackage{commath}
\usepackage{caption}
\usepackage{graphicx}
\usepackage{subcaption}
\usepackage{amsmath, amsfonts, amsthm}
\usepackage{float}
\usepackage{booktabs,siunitx}
\usepackage{url}
\usepackage{multirow}
\usepackage{amssymb}
\usepackage{mathrsfs}
\usepackage{xcolor}
\usepackage{tikz}
\usetikzlibrary{trees}
\usetikzlibrary{shapes,decorations,arrows,calc,arrows.meta,fit,positioning}
\usetikzlibrary{shapes.multipart}
\tikzset{
    -Latex,auto,node distance =1 cm and 1 cm,semithick,
    state/.style ={ellipse, draw, minimum width = 0.7 cm},
    state1/.style ={ draw, minimum width = 0.7 cm},
    point/.style = {circle, draw, inner sep=0.04cm,fill,node contents={}},
    bidirected/.style={Latex-Latex,dashed},
    el/.style = {inner sep=2pt, align=left, sloped}
}
\usepackage[margin=1in]{geometry}

\usepackage{booktabs,siunitx}
\usepackage{amssymb}
\usepackage{authblk}

\usepackage{titlesec}
\titleformat{\paragraph}
{\normalfont\normalsize\bfseries}{\theparagraph}{1em}{}
\titlespacing*{\paragraph}
{0pt}{3.25ex plus 1ex minus .2ex}{1.5ex plus .2ex}

\usepackage{listings}
\allowdisplaybreaks
\usepackage{bbm}
\usepackage{textcomp}
\usepackage[margin=1in]{geometry}
\usepackage[ruled]{algorithm2e}
\SetKwInput{KwParam}{Parameter}
\SetAlgoCaptionLayout{centerline}

\usepackage{sectsty}
\usepackage[onehalfspacing]{setspace}
\usepackage[utf8]{inputenc}

\def\E{{\mathrm E}}
\def\R{\mathbbm{R}}

\def\anhc{\widehat\bmtheta_{{}_{\mathrm{ANHC}}}}
\def\an{\widehat\bmtheta_{{}_{\mathrm{AN}}}}
\def\anc{\widehat\bmtheta_{{}_{\mathrm{ANC}}}}
\def\var{{\mathrm{var}}}
\def\cov{{\mathrm{cov}}}

\def\bmA{\bm A}
\def\bmB{\bm B}
\def\bma{\bm a}
\def\bmx{\bm x}

\def\bmc{\bm c}
\def\impx{\bmX^{\mathrm{imp}}}

\def\bimpx{\bmX^{\mathrm{imp}}}
\def\barimpx{\overline{\bmX}^{\mathrm{imp}}}

\def\bbeta{\bm \beta}

\def\bmX{\bm X}
\def\bmR{\bm R}
\def\bmtheta{\bm \theta}
\def\bmbeta{\bm \beta}
\def\bgamma{\bm \gamma}
\def\bb{\bm b}
\def\br{\bm r}
\newtheorem{condition}{Condition}

\newtheorem{theorem}{Theorem}

\newtheorem{corollary}{Corollary}

\newtheorem{lemma}{Lemma}

\def\anhct{\widehat\bmtheta_{{}_{\mathrm{ANHC}}}}

\def\anct{\widehat\bmtheta_{{}_{\mathrm{ANC}}}}

\def\anhcs{\sigma_{{}_{\mathrm{ANHC}}}}
\def\ans{\sigma_{{}_{\mathrm{AN}}}}

\DeclareMathOperator*{\argmin}{arg\,min}

\allowdisplaybreaks

\begin{document}


\title{
	Adjusting for Incomplete Baseline Covariates in Randomized Controlled Trials: A Cross-World Imputation Framework}
\date{\vspace{-5ex}}

\author{Yilin Song}
\author{James P. Hughes}
\author{Ting Ye}

\affil{Department of Biostatistics, University of Washington,  Seattle, Washington 98195, U.S.A.\thanks{tingye1@uw.edu. This work was supported by the HIV Prevention Trials Network (HPTN) and NIH grant: NIAID 5 UM1 AI068617.}}

\maketitle

\begin{abstract}
\input{sections_arxiv/00_abstract}
\end{abstract}

%


\maketitle


%
\newpage
\input{sections_arxiv/01_intro}
\input{sections_arxiv/02_review}
\input{sections_arxiv/03_CWI}

\input{sections_arxiv/04_optimalCWI}

\input{sections_arxiv/05_simulation}
\input{sections_arxiv/06_realdata}
\input{sections_arxiv/07_discussion}

\section*{Acknowledgements}
\input{sections_arxiv/08_ac}


\section*{Supplementary material}
Web Appendices referenced in Sections 2-6 are available with this paper at the Biometrics website on Oxford Academic. Code necessary to reproduce simulation and application results can be found in the supplementary materials.


\input{sections_arxiv/10_dataavailability}


\newpage
\bibliographystyle{biom}  
\bibliography{reference.bib}

\newpage
\begin{center}
{\sffamily\bfseries\LARGE
Supplementary Materials
}
\end{center}
\setcounter{equation}{0}
\setcounter{table}{0}
\setcounter{lemma}{0}
\setcounter{section}{0}
\renewcommand{\theequation}{S\arabic{equation}}
\renewcommand{\thetable}{S\arabic{table}}
\renewcommand{\thefigure}{S\arabic{figure}}
\renewcommand{\thesection}{S\arabic{section}}
\renewcommand{\thelemma}{S\arabic{lemma}}

\begin{abstract}
Section S1 contains additional theoretical results, and Section S2 contains technical proofs.  
\end{abstract}

\input{sections_arxiv/09_supp}

\end{document}

%% file: sections_arxiv/00_abstract.tex
In randomized controlled trials, adjusting for baseline covariates is commonly used to improve the precision of treatment effect estimation. However, covariates often have missing values. Recently, \cite{zhao2022adjust} studied two simple strategies, the single imputation method and missingness indicator method (MIM), to handle missing covariates and showed that both methods can provide an efficiency gain compared to not adjusting for covariates. To better understand and compare these two strategies, we propose and investigate a novel theoretical imputation framework termed cross-world imputation (CWI). This framework includes both single imputation and MIM as special cases, facilitating the comparison of their efficiency. Through the lens of CWI, we show that MIM implicitly searches for the optimal CWI values and thus achieves optimal efficiency. We also derive conditions under which the single imputation method, by searching for the optimal single imputation values, can achieve the same efficiency as the MIM.
We illustrate our findings through simulation studies and a real data analysis based on the Childhood Adenotonsillectomy Trial. We conclude by discussing the practical implications of our findings.

%% file: sections_arxiv/01_intro.tex
\section{Introduction} 
\subsection{Missing covariates values in randomized controlled trials}
Randomized controlled trials (RCTs) are the gold standard for evaluating treatment effects. Because randomization ensures the treatment groups are comparable at the baseline, a simple comparison of the outcome means across treatment groups is an unbiased estimator of the treatment effect. To improve efficiency, it is common to adjust for baseline covariates in estimating the treatment effects \citep{Fisher:1935aa, fda:2019aa}, {and many have recommended an ANalysis of HEterogeneous COVAriance (ANHECOVA) estimator 
obtained from fitting a linear working model with treatment indicators, }centered covariates, and their interactions as it can achieve guaranteed efficiency gain over the unadjusted estimator \citep[among others]{Tsiatis:2008aa, Lin:2013aa}; in fact, it achieves optimal efficiency among a class of linearly-adjusted estimators \citep{ye2020principles}.

    
Missingness in covariates is common in RCTs, which raises an important question: how should we adjust for covariates with missing values? One intuitive strategy is the \textit{single imputation method}, which imputes all missing values based on the observed data and then adjusts for the imputed covariates. A common choice is to impute by the covariate-wise observed means \citep{whiteandthompson, sullivanandwhite}. Another well-known strategy is the \textit{missingness-indicator method} (MIM), which imputes the missing covariates with zeros and includes the missingness indicators as part of the covariate vector for adjustment \citep{cohen1975applied, rosenbaumandrubin}. Under a randomization inference framework, \citet{zhao2022adjust} showed that MIM used in combination with ANHECOVA is an appealing strategy in RCTs because it does not require modeling the missingness mechanism and is fully robust as long as the missingness is balanced by randomization. It also guarantees efficiency gain compared to the complete-covariate analysis when the missing covariates or missingness indicators are prognostic to the outcome. Furthermore, it is asymptotically no less efficient than the single imputation method. MIM has also been used in combination with weighting estimators, which is asymptotically equivalent to the ANHECOVA counterpart \citep{chang2022covariate}. Note that these appealing properties are due to randomization in RCTs and do not hold in observational studies; see caveats of MIM when used in observational studies in {\cite{greenland1995critical} and \cite{yangandwangandding}.} 

Although MIM is asymptotically non-inferior to the single imputation method,  the efficiency gain may not be large if the missing rate is low (as is common in RCTs) or the missingness indicator is not very prognostic to the outcome. Moreover, MIM needs to adjust for all the missing indicators, which may not be feasible with a small to moderate sample size. In these situations, the single imputation method that uses a more parsimonious model may still be preferable. Therefore, it is of interest to get a more detailed characterization of the efficiency comparison between the single imputation method and MIM, and investigate ways to improve the efficiency of the single imputation method by optimizing over the imputation values. These questions have not been studied in the literature. The main challenge is that with single imputation, the resulting estimator's asymptotic variance is non-convex in the imputation values, complicating both the characterization of the optimal single  imputation  method and its comparison with MIM. 

To address this challenge and bridge this crucial gap, we take a different approach and propose  a novel theoretical  imputation framework termed cross-world imputation (CWI), which includes single imputation and MIM {as special cases and facilitates the efficiency comparison between them.} Using the CWI framework, we show that MIM implicitly searches for the optimal CWI values and thus achieves optimal efficiency in the asymptotic sense. We also derive conditions under which the single imputation method, by searching for the optimal single imputation values, can achieve the same asymptotic efficiency as the MIM.  Based on these theoretical findings, we offer practical recommendations for choosing methods to handle missing covariate values. 


\subsection{Example: Childhood Adenotonsillectomy Trial (CHAT)}

The Childhood Adenotonsillectomy Trial (CHAT) is a multicenter, single-blind, randomized controlled trial 
to evaluate the efficacy of early adenotonsillectomy compared to watchful waiting with supportive care in children with obstructive sleep apnea syndrome \citep{marcuschat,zhangchat}. The trial randomly assigned 464 children aged 5 to 9 years. The primary outcome is the change in the attention and executive-function score on the Developmental Neuropsychological Assessment, which ranges from 50 to 150, with higher values indicating better functioning.


To improve the efficiency of estimating the treatment effect, we consider ANHECOVA with adjustments for prognostic baseline covariates, including baseline body mass index (BMI), gender, tonsil size, Pediatric Sleep Questionnaire (PSQ) summary score, and the 18-item quality-of-life instrument for obstructive sleep apnea (OSA-18) summary score. Despite the relatively low missing rates of these baseline variables (0\% - 3.3\%), challenges remain in determining the best way to adjust for these covariates with missing values, such as deciding which method to use to handle the missing covariate values and how many covariates to adjust for.

The rest of the article proceeds as follows. After introducing the notation and setup, and reviewing existing covariate adjustment methods in Section \ref{sec: review of covariate adjustment methods}, we present the CWI framework and establish its asymptotic properties in Section \ref{method:cwi}. In Section \ref{sec: optimal CWI}, we show that MIM and the single imputation are special cases of CWI and discuss when they achieve optimal efficiency based on the CWI framework. We illustrate our findings by simulations and a real data analysis in Section \ref{sec: sim} and \ref{sec: real data}. We conclude with a discussion in Section \ref{sec: discussion}.

%% file: sections_arxiv/02_review.tex

\section{Review of covariate adjustment methods}
\label{sec: review of covariate adjustment methods}

We consider a $k$-arm trial with pre-specified {expected} treatment assignment proportions $\pi_1,  \dots, \pi_k$, where $0<\pi_t<1$ and $\sum_{t=1}^k \pi_t = 1 $. Let $\bmA_i$ be the $k$-dimensional treatment indicator vector that equals $\bma_t$ if patient $i$ receives treatment $t$, where $\bma_t$ denotes a $k$-dimensional column vector whose $t$th component is 1 and other components are 0. Let $Y_i^{(t)}$ be the potential outcome for patient $i$ under treatment $t$, $t=1,\dots, k$, and $\bmtheta$ be the $k$-dimensional vector whose $t$th component is $\theta_t= E(Y_i^{(t)})$, the population mean of potential (continuous or discrete) outcome under treatment $t$. We are interested in  $\bmtheta$ and given contrasts of $\bmtheta$.  The observed outcome is $Y_i= Y_i^{(t)}$ if and only if $\bmA_i=\bma_t$. For patient $i$, we use  $ \bmX_i = (X_{i1}, \dots, X_{iJ})^\top $ to denote the $J$-dimensional covariate vector for adjustment and use $\bmR_i =(R_{i1}, \dots, R_{iJ})^\top  $ to indicate missingness of covariates, where $R_{ij}=1$ if $X_{ij}$ is observed and $R_{ij}=0$ if $X_{ij}$ is missing. We consider simple randomization where {$\bmA_i$ is independent of $ (Y_i^{(1)},\dots, Y_i^{(k)}, \bmR_i, \bmX_i )$ for each $i$ } and $P(\bmA_i=\bma_t)=\pi_t$ for $t=1\dots, k$. We also impose the following mild condition.
\begin{condition}\label{cond: iid}
	$ (Y_i^{(1)},\dots, Y_i^{(k)}, \bmR_i, \bmX_i ), i=1,\dots, n$ are independent and identically distributed with finite second order moments. The distribution of baseline covariates $\bmX_i$ and missing indicators  $\bmR_i$ are not affected by treatment. 
\end{condition} 
Condition \ref{cond: iid} asserts the missingness is unaffected by the treatment assignment, which holds if $\bmX_i$ is measured at baseline so the covariates and their
missingness indicators in different arms are balanced by randomization. To simplify the notation, {we drop the subscript $i$ for a generic unit.}

\subsection{Covariate adjustment with completely observed covariates}\label{sec:review} 

Because of randomization, the simplest estimator of $\bmtheta$ is the ANalysis Of VAriance (ANOVA) estimator $\an= (\overline Y_1, \dots, \overline Y_k)^\top$, where $\overline Y_t$ is the sample average of the outcomes under each treatment $t$. To adjust for  completely observed covariates, two \textit{working models} are commonly used. These models are not required to be correct; they are only used as working models to obtain the corresponding estimators. 

The first model is ANalysis of COVAriance (ANCOVA) \citep{Fisher:1935aa}{{, written using the \textsf{R} convention}}: $Y\sim \bmA+ (\bmX - \overline \bmX) $, where $\overline \bmX$ is the sample average of $\bmX_i$'s. The coefficient vector of $\bmA$ is a consistent estimator of $\bmtheta$ regardless of whether the model is correct or not. However, when the model is incorrect, the ANCOVA estimator $\anc$ can be even less efficient than the simple ANOVA estimator \citep{Freedman:2008aa}.

The second model is ANalysis of HEterogeneous COVAriance which further includes the treatment-by-covariate interactions in the model \citep{Yang:2001aa, Tsiatis:2008aa, Lin:2013aa, ye2020principles}: $Y\sim \bmA+\left(\bmX-\overline\bmX\right) I(\bmA= \bma_1) +\dots + \left(\bmX-\overline\bmX\right) I(\bmA= \bma_k)$. Similar to ANCOVA, the coefficient vector of $\bmA$ is a consistent estimator of $\bmtheta$ regardless of whether the model is correct or not. Moreover, even when the model is incorrect, the ANHECOVA estimator $\anhc$ is never less efficient than the ANCOVA estimator $\anc$ and the ANOVA estimator $\an$ \citep{ye2020principles}.

Note that centering the covariates using $\bmX-\overline \bmX$ is important for $\anc$ and $\anhc$ to consistently estimate $\bmtheta$. The only exception is when ANCOVA is used to estimate linear contrasts of $\bmtheta$ such as $\theta_t-\theta_s$ for $t,s =1, \dots, k$, when the covariate mean $\overline \bmX$ gets canceled.

\subsection{Covariate adjustment with missing covariates}
\label{subsec: missing} 

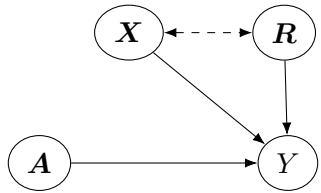
\begin{figure}[H]
	\centering
	\tikzset{
		-Latex,auto,node distance =1 cm and 1 cm,semithick,
		state/.style ={ellipse, draw, minimum width = 0.7 cm},
		state1/.style ={ draw, minimum width = 0.7 cm},
		point/.style = {circle, draw, inner sep=0.04cm,fill,node contents={}},
		bidirected/.style={Latex-Latex,dashed},
		el/.style = {inner sep=2pt, align=left, sloped}
	}
	\centering
	\resizebox{.25\textwidth}{!}{\begin{tikzpicture}
			\node[state] (a) at (0,0) {${\bmA}$};
			\node[state] (y) [right =of a, xshift=1.5cm] {${Y}$};
			\node[state] (r) [above =of y, xshift=-0.05cm] {$\bmR$};
			\node[state] (x) [above =of a, xshift=1.2cm, yshift=0cm] {$\bmX$};
			\path (a) edge node[above]{} (y);
			\path (r) edge node[above]{} (y);
			\path[bidirected] (x) edge node[above]{} (r);
			\path (x) edge node[above]{} (y);
	\end{tikzpicture}} 
	\caption{A directed acyclic graph (DAG) of the missingness assumption and simple randomization.}
	\label{fig: missing}
\end{figure}

As depicted in Figure \ref{fig: missing}, the key missing data assumption in Condition \ref{cond: iid} is that the treatment indicators are independent of the covariates $\bmX$ and the missingness indicators $\bmR$, which holds if the covariates are measured at the baseline, and thus the distribution of   $\bmX$ and $\bmR$ is balanced by randomization. Meanwhile, the relationship among $\bmX, \bmR, Y$ can be complex, e.g., the missingness can depend on the covariates $\bmX$ {and the potential outcomes $Y^{(1)},\dots, Y^{(k)}$}, which, following \cite{rubin1976inference}, is known as missing not at random. The act of randomization obviates the need to model the missingness mechanism and allows for robust estimation of the treatment effect when using the partially observed covariate information. 

We review two popular methods below. The first method is the \textit{single imputation} (SI) method, which imputes the missing covariates once with some fixed or data-dependent values and then adjusts for the imputed covariates.  \cite{zhao2022adjust} considered a covariate-wise imputation scheme that imputes all missing values of the $j$th covariate $X_{ij}$ by a fixed or data-dependent constant $\widehat{c}_j$. For example, it is common to impute by the observed covariate-wise average $ \widehat c_j = \sum_{i=1}^{n} R_{ij} X_{ij}/ \sum_{i=1}^{n} R_{ij} $. For the $i$th patient, write the imputed $X_{ij}$ as $X_{ij}^{\mathrm{imp}} (\widehat c_j)=  R_{ij} X_{ij} + (1-R_{ij}) \widehat c_j$ and the imputed covariate vector as $ \bimpx_i (\widehat \bmc) = \{X_{i1}^{\mathrm{imp}}(\widehat c_1), \dots, X_{iJ}^{\mathrm{imp}} (\widehat c_J) \}^\top $, which can also be expressed as $ \bimpx_i (\widehat \bmc) = \bmR_i \circ \bmX_i +  (\bm 1_J -\bmR_i) \circ \widehat \bmc$, where $\bmR_i \circ \bmX_i = (R_{i1}X_{i1}, \dots,  R_{iJ}X_{iJ})^\top $ and $(\bm 1_J - \bmR_i) \circ  \widehat \bmc = \{(1-R_{i1})\widehat c_1, \dots, (1-R_{iJ})\widehat  c_J\}^\top $  with $\circ$ representing the Hadamard product, $\widehat\bmc= (\widehat c_1, \dots, \widehat c_J)^\top$,  and $\bm 1_J$ a $J$-dimensional column vector of 1's.  Then, we can proceed with fitting ANCOVA or ANHECOVA working model and obtain the covariate-adjusted estimators $ \anct^{\mathrm{si}} (\widehat \bmc)$ and $ \anhct^{\mathrm{si}} (\widehat \bmc)$. These estimators are fully robust in the sense that they are consistent to $\bmtheta$ even when the missingness depends on the underlying covariates themselves and the potential outcomes. For two treatment arms and estimating the average treatment effect $\theta_2 - \theta_1$, \cite{zhao2022adjust} studied the SI strategy under a randomization inference framework and proved that the resulting ANHECOVA estimator $ \anhct^{\mathrm{si}} (\widehat \bmc)$ is always more efficient than the unadjusted estimator $\an$. 

The second method is MIM, which includes $\bmR$ as additional regressors in the model. For example, using either the ANCOVA or ANHECOVA working model 
\begin{align*}
   &  Y \sim  \bmA+\{\impx(\widehat\bmc)-\barimpx(\widehat\bmc)\}+\left(\bmR-\overline\bmR\right)\\
   &  Y \sim  \bmA+I(\bmA=\bma_1)\{\impx(\widehat\bmc)-\barimpx(\widehat\bmc)\}+\dots +I(\bmA=\bma_k)\{\impx(\widehat\bmc)-\barimpx(\widehat\bmc)\}\\
    &\quad\quad  +I(\bmA=\bma_1)\left(\bmR-\overline\bmR\right) + \dots + I(\bmA=\bma_k)\left(\bmR-\overline\bmR\right),
\end{align*}  
where $\overline \bmR$ is the sample average of $\bmR_i$'s,  the obtained coefficient vector of $\bmA$ is a consistent estimator of $\bmtheta$. Compared to the SI, the main advantage of MIM is that including $\bmR$ as regressors removes the dependence of the resulting estimators on the imputation values $\widehat\bmc$. For two arms and estimating the average treatment effect $\theta_2 - \theta_1$, \citet{zhao2022adjust} proved that the ANHECOVA estimator combined with MIM can further improve the efficiency over the SI counterpart $ \anhct^{\mathrm{si}} (\widehat \bmc)$ if $\bmR$ is related to the potential outcomes. 

Strictly speaking, if a covariate $X_j$ is observed for all subjects, then $R_j=1$ for all units and is not needed to be adjusted in the model. Also, to prevent colinearity, if the missingness indicator $R_t=R_s$ for covariate $X_t, X_s$, only one of the missingness indicators need to be adjusted. From now on, we will use $\bmR$ to denote the vector of missingness indicators included in the model after appropriate adjustment for simplicity of notation.

As mentioned in \cite{zhao2022adjust}, MIM is not a new method \citep{whiteandthompson, carpenter, Rubinbook, rodericklittle, robinsetal, jones, ibrihimetal}. The use of MIM in observational studies has been criticized by \cite{greenland1995critical}, \cite{donders2006gentle} and \cite{yangandwangandding}. However, \cite{zhao2022adjust} showed that MIM is a robust and efficient method when used in RCTs.


%% file: sections_arxiv/03_CWI.tex

\section{A cross-world imputation (CWI) framework}\label{method:cwi}

We propose a novel theoretical imputation framework called the \textit{cross-world imputation}, which includes the methods reviewed in Section \ref{subsec: missing} as special cases and facilitates the efficiency comparison between the reviewed strategies. We name this procedure CWI because it allows for different imputation values when estimating different potential outcome means. 
 For the purpose of this article, we focus on linear adjustment; we leave non-linear adjustment as future work. For CWI, we only consider adjusting for the imputed $\bmX$, but not the missingness indicators $\bmR$. 
 
 For treatment arm $t, t=1,\dots, k$,  the steps for CWI are as follows: 
\begin{enumerate}
    \item  Let $\widehat{c}_{tj}$ be a pre-specified fixed or data-dependent imputation value for the $j$th covariate, and $\widehat \bmc_t = (\widehat c_{t1}, \dots, \widehat c_{tJ})^\top $ be the vector of imputation values for all covariates. Apply covariate-wise SI with $\widehat \bmc_t $, i.e., impute $ \widehat c_{tj}$ for all the missing values in the $j$th covariate and denote the imputed covariate as $\impx (\widehat \bmc_t)$. 
    \item Fit a model $Y\sim 1+ \impx (\widehat \bmc_t) $ using data with $\bmA=\bma_t$ and constrained least squares 
    $$
    (\widehat \alpha_t, \widehat\bbeta_t)= \argmin_{\alpha \in \mathbb{R}, \bbeta \in \mathcal{R}_{\bbeta}}
    \sum_{i: \bmA_i = \bma_t} \big[Y_i - \alpha - \bbeta^\top \{\impx_i (\widehat \bmc_t) - \overline{\bmX}^{\mathrm{imp}}_t (\widehat \bmc_t)\} \big]^2,
    $$
    where 
    $\mathcal{R}_{\bbeta}$ can be any constrained set or the unconstrained $\mathbb{R}^{J}$. The resulting model {gives} $\widehat{\mu}_t(\bmx;\bb;{\widehat \bmc_t})= \overline Y_t + \bb^\top \{\bmx -  \overline{\bmX}^{\mathrm{imp}}_t (\widehat \bmc_t)\}$, for some $\bb$. 
    \item Compute  the predicted values for all subjects, including those not in arm $t$. Their average
    $$
    \widehat\theta^{\mathrm{cwi}}_t (\bb; \widehat \bmc_t)= n^{-1}\sum_{i=1}^n \widehat{\mu}_t (\impx_i (\widehat \bmc_t); \bb; {\widehat \bmc_t}) = \overline Y_t - \bb^\top \{\overline{\bmX}^{\mathrm{imp}}_t (\widehat \bmc_t) - \overline{\bmX}^{\mathrm{imp}} (\widehat \bmc_t) \} 
    $$ 
    is an estimator of $\theta_t=E(Y^{(t)})$.
\end{enumerate} 

\begin{figure}[H]
	\centering
	\includegraphics[width=\linewidth]{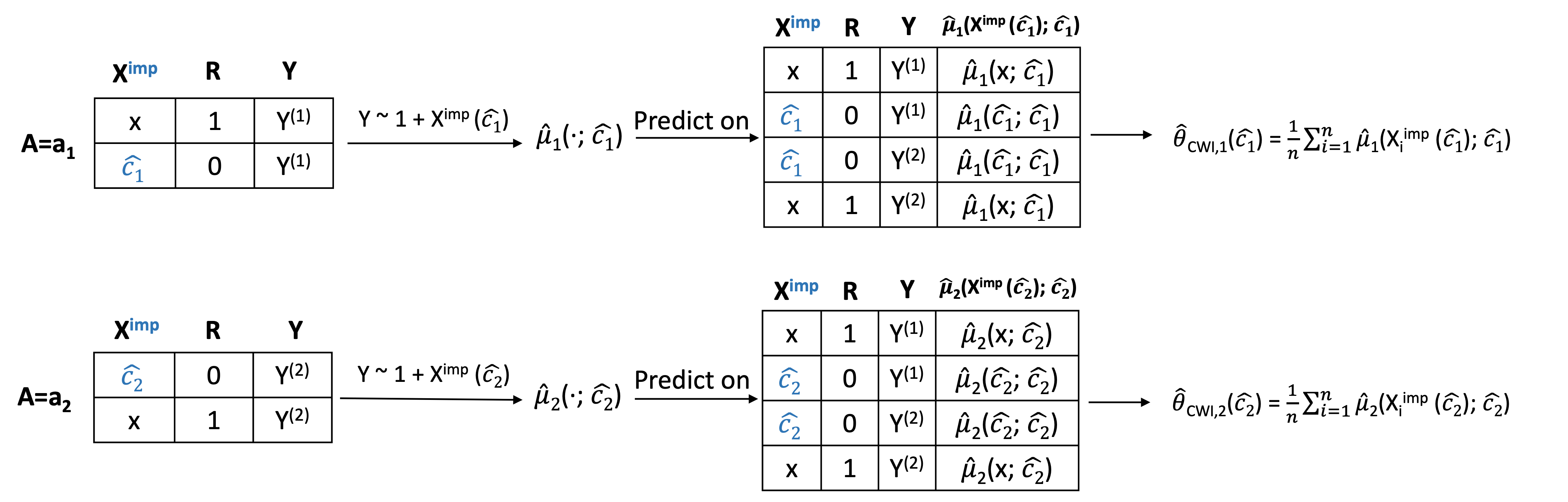}
	\caption{An illustration of CWI with two treatment arms and a single covariate. \textcolor{black}{This is a simplified example that we only have two treatment arms and one baseline covariate $X$, and each arm has two observations including one with missing baseline covariate. We first impute $\widehat{c}_1$ for the missing value in the covariate to obtain the imputated covariate $X^{\mathrm{imp}}(\widehat{c}_1)$. Then, we fit a model $Y\sim 1+ \impx (\widehat c_1) $ using data with $\bmA=\bma_1$. Using the resulting model, we compute the predicted values for all subjects with $\widehat{c}_1$ imputed for all missing values in the covariate in both arms. The mean of predicted values of all subjects is an estimator of $\theta_1$. We repeat the same steps with $\widehat{c}_2$ and treatment arm $\bmA=\bma_2$ to obtain an estimator of $\theta_2$. }}
	\label{fig:imp}
\end{figure}

An illustration of CWI with two treatment arms and a single covariate is in Figure \ref{fig:imp}. The resulting estimator is termed \textit{g-computation} or \textit{standardization} estimator when there are no missing covariates \citep{freedman2008randomization}. Under CWI, an important example is when $\mathcal{R}_{\bbeta} = \mathbb{R}^J$, i.e., when the fitted model $\widehat\mu_t$ is from the unconstrained least squares, the resulting estimator is the ANHECOVA estimator under CWI, denoted as $\widehat\bmtheta_{{}_{\mathrm{ANHC},t}}^{\mathrm{cwi}} (\widehat \bmc_t)$. When there are no missing covariate values,  $\widehat\bmtheta_{{}_{\mathrm{ANHC},t}}^{\mathrm{cwi}} (\widehat \bmc_t)$ obtained from the g-computation formulation (Steps 1-3) is the same as the $\widehat\bmtheta_{{}_{\mathrm{ANHC},t}} $ obtained from the regression model discussed in  Section \ref{sec:review}. However, when there are missing covariate values, the g-computation formulation gives us more flexibility in the imputation scheme compared to the regression formulation: the g-computation formulation allows CWI whereas the regression formulation does not.  

We remark that one can also obtain an ANCOVA-type estimator under CWI by constraining the coefficients of $\impx(\widehat\bmc_t)$ being equal across $t$ when performing the constrained ordinary least squares in Step 2. However, we do not consider it further because its computation is complex and,  as will be shown in Section \ref{subsec: optimality mim}-\ref{subsec: efficiency between mim and si}, it is less efficient than ANHECOVA. 

It is straightforward to see that CWI includes SI as a special case. Specifically, when  $\widehat \bmc_1=\dots= \widehat \bmc_k$, CWI reduces to SI. However, CWI is distinct from imputing by treatment-specific values in that CWI imputes all subjects $k$ times, respectively with $\widehat \bmc_1, \dots,  \widehat \bmc_k$ for estimating the response means in $k$ arms, while imputing by treatment-specific values only imputes all subjects once -- those with $\bmA=\bma_t$ are imputed by $\widehat\bmc_t$. This distinction has an important implication: the probability limit of $\widehat \bmc_t$ being dependent on $t$ does not bias CWI but can bias imputation by treatment-specific values. In Lemma S1 of the Supplementary Materials, we show that imputation by arm-specific observed covariate mean leads to a less efficient ANHECOVA estimator  compared to imputation by overall observed covariate mean. Hence, we will not consider imputation by arm-specific values further beyond this point. Though not obvious at this point, we will show in Section \ref{sec:equi} that CWI also includes MIM as a special case.

Now, to unify the notation going forward, we use $\widehat{\bmtheta}(\widehat{\bb}_1, \dots, \widehat \bb_k; \widehat{\bmc}_1, \dots, \widehat{\bmc}_k)$ to define a general class of estimators  for potential outcome means in $k$ treatment arms under CWI:
\begin{align}
	\widehat{\bmtheta}(\widehat{\bb}_1, \dots, \widehat \bb_k; \widehat{\bmc}_1, \dots, \widehat{\bmc}_k) = \left[\overline{Y}_t-\widehat{\bb}_t^\top\big\{\overline{\bmX}^{\mathrm{imp}}_t(\widehat{\bmc}_t)-\overline{\bmX}^{\mathrm{imp}}(\widehat{\bmc}_t)\big\}, t=1,\dots, k\right]^\top,
	\label{def:theta}
\end{align}
where $\widehat{\bb}_t$ can be fixed or stochastic. In the notation below, we specify the imputation method (SI, CWI) as superscripts and the estimator type (ANOVA(AN), ANCOVA(ANC), ANHECOVA(ANHC)) as subscripts. The general class of estimators in \eqref{def:theta} includes all the
aforementioned estimators under SI or CWI:
\begin{align*}
	\widehat{\bmtheta}(\widehat{\bb}_1, \dots, \widehat \bb_k; \widehat{\bmc}_1, \dots, \widehat{\bmc}_k)= \begin{cases}\an & \text { if } \widehat{\bb}_t=0 \text { for all } t \\ 
		\anc^{\mathrm{si}} & \text { if } \widehat{\bmc}_1=\cdots=\widehat{\bmc}_k=\widehat{\bmc}, \text{ and } \widehat{\bb}_t=\widehat{\bbeta}(\widehat{\bmc}) \text { for all } t\\
		\anhc^{\mathrm{si}} & \text { if } \widehat{\bmc}_1=\cdots=\widehat{\bmc}_k=\widehat{\bmc}, \text{ and } \widehat{\bb}_t=\widehat{\bbeta}_t(\widehat{\bmc}) \text { for all } t\\
		\anhc^{\mathrm{cwi}} & \text { if }  \widehat{\bb}_t=\widehat{\bbeta}_t(\widehat{\bmc}_t) \text { for all } t 
	\end{cases} 
\end{align*}
where 
{\small
\begin{align*}
	\widehat{\bbeta}(\widehat{\bmc}) & =\bigg[\sum_{t=1}^k \sum_{i: \bmA_i=\bma_t}\big\{\impx_i(\widehat\bmc)-\overline{\bmX}_t^{\mathrm{ imp}}(\widehat{\bmc})\big\}\big\{\impx_i(\widehat\bmc)-\overline{\bmX}^{\mathrm{ imp}}_t(\widehat{\bmc})\big\}^\top\bigg]^{-1} \sum_{t=1}^k \sum_{i: \bmA_i=\bma_t}\big\{\impx_i(\widehat\bmc)-\overline{\bmX}_t^{\mathrm{imp}}(\widehat{\bmc})\big)\}Y_i\\
	\widehat{\bbeta}_t(\widehat{\bmc}_t) & = \bigg[\sum_{i: \bmA_i=\bma_t}\big\{\impx_i(\widehat\bmc_t)-\overline{\bmX}_t^{\mathrm{imp}}(\widehat{\bmc}_t)\big\}\big\{\impx_i(\widehat\bmc_t)-\overline{\bmX}^{\mathrm{imp}}_t(\widehat{\bmc}_t)\big\}^\top\bigg]^{-1} \sum_{i: \bmA_i=\bma_t}\big\{\impx_i(\widehat\bmc_t)-\overline{\bmX}_t^{\mathrm{imp}}(\widehat{\bmc}_t)\big\} Y_i.
\end{align*}}

Theorem \ref{theo:asymp} gives the asymptotic property for the general class of estimators in \eqref{def:theta}.
\begin{theorem}\label{theo:asymp}
	Suppose that  $ \widehat \bmc_t = (\widehat c_{t1},\dots, \widehat c_{tJ})^\top$ has a finite probability limit $ \bmc_t =(c_{t1},\dots, c_{tJ})^\top $, i.e., $ \widehat \bmc_t \xrightarrow{p} \bmc_t $, and  $\widehat{{\bb}}_t\xrightarrow{p} \bb_t$  as $n\to \infty$, for $t=1,\dots, k$. Also suppose that $\var(\bimpx (\bmc_t)) $ is positive definite. Under Condition \ref{cond: iid} and simple randomization,  as $n\to \infty$, 	$\sqrt{n}\Big(  \widehat{\bmtheta}(\widehat{\bb}_1, \dots, \widehat \bb_k; \widehat{\bmc}_1, \dots, \widehat{\bmc}_k)-\bmtheta\Big)$ is asymptotically normal with mean 0 and variance $V(\bmB; \bmc_1, \dots, \bmc_k )$. The expression of \\ $V(\bmB; \bmc_1, \dots, \bmc_k )$ is in Section S1 of the Supplementary Materials.
\end{theorem}

The proof of Theorem \ref{theo:asymp} is in the Supplementary Materials. In the proof, we note that for the general class of CWI estimators,  $\widehat{\bmtheta}(\widehat{\bb}_1, \dots, \widehat \bb_k; \widehat{\bmc}_1, \dots, \widehat{\bmc}_k)=\widehat{\bmtheta}({\bb}_1, \dots,  \bb_k; {\bmc}_1, \dots, {\bmc}_k)+o_p(1/\sqrt{n})$, which implies that CWI with data-dependent imputation values $\widehat \bmc_1,\dots, \widehat\bmc_k$ is asymptotic equivalent with imputing using their probability limits $ \bmc_1,\dots,\bmc_k$. As linear contrasts of outcome means in two arms are often of interest in many clinical trials, in Corollary S1 of the Supplementary Materials, we provide the asymptotic variance of linear contrasts using ANHECOVA. 

Theorem \ref{theo:asymp} also reiterates that the estimators considered in \eqref{def:theta} are fully robust in the sense that they remain consistent to $\bmtheta$ even when the missingness depends on the underlying covariates and the potential outcomes, or when the linear working models used are misspecified.

%% file: sections_arxiv/04_optimalCWI.tex

\section{Optimal cross-world imputation}\label{sec: optimal CWI}

{In this section, we will use the CWI framework to compare the efficiency between SI and MIM. 

\subsection{MIM as a special case of CWI}\label{sec:equi}

First, we will show that the MIM reviewed in Section \ref{subsec: missing} can be viewed as a special case of CWI with a particular set of imputation values.  

As MIM is invariant to the values used in SI  \citep[Lemma 3]{zhao2022adjust}, {it suffices to consider MIM with $\widehat \bmc = \bm{0}$, so the ANHECOVA estimator with MIM can be written as {$ \anhc^{{\mathrm{mim}}} (\bm{0}) = \{\widehat\theta_{{}_{{\mathrm{ANHC}},1}}^{{\mathrm{mim}}} (\bm{0}), \dots, \widehat\theta_{{}_{{\mathrm{ANHC}},k}}^{{\mathrm{mim}}} (\bm{0})\}^\top$,} where 
$
 \widehat\theta_{{}_{{\mathrm{ANHC}},t}}^{{\mathrm{mim}}} (\bm{0})  =   \overline Y_t  - \widetilde \bbeta_{t}(\bm{0})^\top \{ \overline{\bmX}_{t}^{\mathrm{imp}} (\bm{0}) -  \overline{\bmX}^{\mathrm{imp}} (\bm{0})\} - \widetilde {\bgamma}_{t}(\bm{0})^\top ( \overline{\bmR}_t -  \overline{\bmR}) ,
$
 $\widetilde \bbeta_{t}(\bm 0)$ and $\widetilde {\bgamma}_{t} (\bm 0)$ are respectively the estimated coefficient vectors of $\impx(\bm 0)$ and $\bmR$ in the linear model $Y\sim 1+ \impx(\bm{0}) + \bmR$ fitted using those with $\bmA_i=\bma_t$. Note that $\impx(\bm{0}) = \bmR \circ \bmX $. 
On the other hand, consider a CWI scheme with $\widehat\bmc_t =  -  \widetilde {\bgamma}_{t}(\bm{0}) \oslash   \widetilde {\bbeta}_{t}(\bm{0}):=\widehat\bmc_t^\star  $, where $\oslash$ denotes Hadamard division. It is interesting to note that the resulting ANHECOVA estimator for $\theta_t$, denoted as $ \widehat\theta_{{}_{{\mathrm{ANHC}},t}}^{{\mathrm{cwi}}} (\widehat\bmc_t^\star )$, is algebraically the same as the MIM estimator $ \widehat\theta_{{}_{{\mathrm{ANHC}},t}}^{{\mathrm{mim}}} (\bm{0}) $. This result is summarized in Lemma \ref{lemma: mim as cwi}. 
\begin{lemma} \label{lemma: mim as cwi}
For $t=1,\dots, k$, when $\widehat \bmc_t^\star= -  \widetilde {\bgamma}_{t}(\bm{0}) \oslash   \widetilde {\bbeta}_{t}(\bm{0}) $ is well-defined, we have  $\widehat\theta_{{}_{{\mathrm {ANHC}},t}}^{{\mathrm{cwi}}} (\widehat\bmc_t^\star )= \widehat\theta_{{}_{{\mathrm {ANHC}},t}}^{{\mathrm {mim}}} (\widehat\bmc_t^\star )= \widehat\theta_{{}_{{\mathrm {ANHC}},t}}^{{\mathrm {mim}}} (\bm{0}) $.
\end{lemma} 
To better understand Lemma \ref{lemma: mim as cwi}, consider a simple situation with a single covariate. MIM with $\widehat c= 0$ is $\widehat\theta_{{}_{{\mathrm {ANHC}},t}}^{{\mathrm {mim}}} (0) =  n^{-1} \sum_{i=1}^n ( \widetilde{\alpha}_t+\widetilde{\beta}_{t} X_i R_i+\widetilde{\gamma}_{t} R_i ) $, where the estimated coefficients are from fitting the linear model $
Y=\widetilde{\alpha}_t+\widetilde{\beta}_{t} XR+\widetilde{\gamma}_{t} R+ \widehat \epsilon
$ using data with $\bmA=\bma_t$. This model can be written as 
\begin{align}
 Y=(\widetilde{\alpha}_t+\widetilde{\gamma}_{t})+\widetilde{\beta}_{t} \underbrace{\big\{ XR+(1-R)(-\widetilde{\gamma}_{t}/\widetilde{\beta}_{t})\big\}}_{X^{\mathrm {imp}} ( -\widetilde{\gamma}_{t}/\widetilde{\beta}_{t})}+ \widehat \epsilon . \label{eq: 1d cwi}
\end{align}
In the proof of Lemma \ref{lemma: mim as cwi}, we show that \eqref{eq: 1d cwi} is the same as the model obtained from fitting $Y\sim 1+ X^{\mathrm {imp}} ( \widehat c_t^
\star) $ using data with $\bmA=\bma_t$, which is the ANHECOVA under CWI with the imputed value $\widehat c_t^
\star= -\widetilde{\gamma}_{t}/\widetilde{\beta}_{t}$.  Hence, $\widehat\theta_{{}_{{\mathrm {ANHC}},t}}^{{\mathrm {cwi}}} (\widehat c_t^
\star) =  n^{-1} \sum_{i=1}^n \{ (\widetilde{\alpha}_t+\widetilde{\gamma}_{t}) +\widetilde \beta_{t} X^{\mathrm {imp}} (\widehat c^\star_t) \} $ is the same as $\widehat\theta_{{}_{{\mathrm {ANHC}},t}}^{{\mathrm {mim}}} (0)$. Another perspective to this result is to start from MIM with $\widehat c =  \widehat c_t^\star$, which equals MIM with $\widehat c = 0$ because MIM is invariant to the imputation value. Then fitting a linear model $Y\sim 1+ X^{\mathrm {imp}} ( \widehat c_t^\star)  + R $ using data with $\bmA=\bma_t$
also results in \eqref{eq: 1d cwi} because $X^{\mathrm {imp}} ( \widehat c_t^\star) $ has effectively summarized the  projection of $Y$ onto the linear space spanned by $XR$ and $R$ and it becomes unnecessary to further adjust for $R$. }

\subsection{MIM achieves optimality among CWI}\label{subsec: optimality mim}

Theorem \ref{theorem:optimality} says that the ANHECOVA estimator with MIM is not only a special case of CWI, it also achieves optimal efficiency among the class of estimators in \eqref{def:theta}. Hence, Theorem \ref{theorem:optimality} provides an interesting perspective that ANHECOVA with MIM actually implicitly searches for the optimal CWI values $\widehat\bmc_t^\star$'s.

\begin{theorem}\label{theorem:optimality}
The $V(\bmB; \bmc_1, \dots, \bmc_k )$ in Theorem \ref{theo:asymp} is minimized at $\anhc^{{\mathrm {mim}}} (\bm{0}) $, the ANHECOVA estimator with MIM. 
\end{theorem} 
The proof of Theorem \ref{theorem:optimality} is in the Supplementary Materials. Here, we outline the key ideas. First, we introduce a larger class of estimators compared to that defined in \eqref{def:theta}: 
{\begin{align}\nonumber
    \widehat{\bmtheta}_{\mathrm L}(\widehat{\bb}_1, \dots, \widehat \bb_k, \widehat{\br}_1,\dots, \widehat{\br}_k; \widehat{\bmc}_1, \dots, \widehat{\bmc}_k)=\left[\overline{Y}_t-\widehat{\bb}_t^\top\{ \overline{\bmX}^{\mathrm {imp}}_t(\widehat{\bmc}_t)-\overline{\bmX}^{\mathrm {imp}}(\widehat{\bmc}_t)\} - \widehat{\br}_t^\top \left(\overline{\bmR}_t-\overline{\bmR}\right), t=1\dots, k \right]^\top. 
\end{align}}
This class includes the MIM used in combination with the CWI, which is the estimator obtained following the procedure in Section \ref{method:cwi} with $Y\sim 1+ \impx (\widehat \bmc_t) + \bmR $ in Step 2. We show that this estimator is invariant to the CWI values $\widehat\bmc_1, \dots, \widehat \bmc_k$, which extends Lemma 3 in  \cite{zhao2022adjust} to the CWI framework. This invariance property implies that even in the CWI framework, there is only one ANHECOVA MIM estimator and thus it suffices to consider $\anhc^{{\mathrm {mim}}} (\bm{0}) $. Next, we show that the class of estimators $\{ \widehat{\bmtheta}_{\mathrm L}({\bb}_1, \dots,  \bb_k, {\br}_1,\dots, {\br}_k; {\bmc}_1, \dots, {\bmc}_k): \text{for any } {\bb}_t, \br_t, {\bmc}_t\}$ is the same as the class $\{ \widehat{\bmtheta}_{\mathrm L}({\bb}_1, \dots,  \bb_k, {\br}_1,\dots, {\br}_k; \bm{0}, \dots, \bm{0}): \text{for any } {\bb}_t, \br_t\}$. Then, Theorem 1 of \cite{ye2020principles} implies that $\anhc^{{\mathrm {mim}}} (\bm{0}) $ is optimal within the latter class in the sense that it has the smallest asymptotic variance. Hence, $\anhc^{{\mathrm {mim}}} (\bm{0}) $ is optimal within the former class as well. Because we showed in Lemma \ref{lemma: mim as cwi} that $\anhc^{{\mathrm {mim}}} (\bm{0}) $ actually belongs to the smaller class $\{ \widehat{\bmtheta}({\bb}_1, \dots,  \bb_k; {\bmc}_1, \dots, {\bmc}_k): \text{for any } {\bb}_t, {\bmc}_t\}$, it must be optimal within this small class, which concludes the proof.

\subsection{When can SI achieve optimality among CWI?} \label{subsec: efficiency between mim and si}

Now that we have embedded both the SI and MIM methods into the CWI framework and showed that MIM achieves optimality, in the following, we will {rely} on the CWI framework to study when SI can also achieve optimality. {Specifically, we are interested in when optimizing over the imputation values for the SI method can achieve optimality among the class of estimators in \eqref{def:theta} and what is the optimal imputation value for SI.} \textcolor{black}{These questions are important because SI is intuitive and easy to implement, and compared to MIM, it adjusts for a smaller number of covariates which is advantageous when we have a small sample size or a large number of covariates with missing values.}
To the best of our knowledge, these questions have not been studied in the literature.

Theorem \ref{lemma:equi} establishes a sufficient condition under which the SI can achieve optimality.
\begin{theorem}\label{lemma:equi}
Under the conditions in Theorem \ref{theo:asymp},  MIM and the optimal SI can achieve the same asymptotic efficiency when the probability limit of
 $\widehat \bmc_t^\star$ does not depend on $t$, which holds when {$\lim_{n\to \infty}\widetilde \gamma_{tj} (\bm{0})/ \lim_{n\to \infty}\widetilde \beta_{tj} (\bm{0})
$ does not depend on $t$ for every $j$.}  
\end{theorem} 

We give the explicit form of $\lim_{n\to \infty}\widetilde \gamma_{tj} (\bm{0})/ \lim_{n\to \infty}\widetilde \beta_{tj} (\bm{0})
	$ in the Supplementary Materials. 
	Theorem \ref{lemma:equi} implies that there are two special situations in which the optimal SI can achieve the same efficiency as MIM, and thus achieve optimality among the class of estimators in \eqref{def:theta}: (i) there are no treatment-by-covariate and treatment-by-missingness interactions in the outcome generating process, i.e., when the probability limits of  $\widetilde{\bgamma}_t (\bm{0})$ and  $\widetilde\bbeta_t(\bm{0})$ do not depend on $t$; {or} (ii) the missingness occurs completely at random, i.e., $\bmR \perp (\bmA, \bmX, Y^{(t)} )$ for all $t$, which implies that $\lim_{n\to\infty}\widetilde{\bgamma}_t (\bm{0})= \bm{0}$ for every $t$.	When the condition in Theorem \ref{lemma:equi} holds, the optimal imputation value for SI is simply the probability limit of
	$\widehat \bmc_t^\star$.	In fact, when the missingness occurs completely at random, the optimal imputation value for SI  (i.e. the probability limit of $\widehat \bmc_t^\star$) is simply a vector of the expected observed covariate means. This implies that SI  by covariate-wise observed means can achieve the same asymptotic efficiency as the MIM under missing completely at random. This result is  summarized  in Corollary \ref{corollary:MCAR}. 
	\begin{corollary}\label{corollary:MCAR}
		When the missingness occurs completely at random, i.e., $\bmR \perp (\bmA, \bmX, Y^{(t)} )$ for all $t$, the probability limit of
		$\widehat \bmc_t^\star$ does not depend on $t$ and equals $\{E(X_{1}\mid R_{1} =1), \dots, E(X_{J}\mid R_{J} =1)\}^\top $. 
	\end{corollary}
	
	Lastly, when the condition in Theorem \ref{lemma:equi} does not hold, the optimal SI is generally less efficient than the MIM in the asymptotically sense, and analyzing the optimal imputation value for SI is complex because the asymptotic variance is not a convex function of $\bmc$. In Section S1 of the Supplementary Materials, we analyze the special case with a single covariate in detail.


%% file: sections_arxiv/05_simulation.tex
\section{Simulation}\label{sec: sim}

We compare the finite sample performance of SI and MIM in simulations with three treatment arms. Our simulation study can be summarized by the factorial design below: 
\begin{description}
	\item[Factor 1] Data generating distribution:  {(Case 1) {Missing completely at random (MCAR):} $\bmR \perp (Y^{(1)}, Y^{(2)}, { Y^{(3)}}, \bmX)$, and $(X_{j}, R_{j}) \perp (X_{k}, R_{k}) $ for $j\neq k$; (Case 2) $\bmR$ depends on $\bmX$, and $(X_{j}, R_{j}) \perp (X_{k}, R_{k}) $ for $j\neq k$; (Case 3) $\bmR$  depends on  $(Y^{(1)}, Y^{(2)}, { Y^{(3)}}, \bmX)$, and $(X_{j}, R_{j}) $ is correlated with $ (X_{k}, R_{k}) $ for $j\neq k$.}
	\item[Factor 2] Sample size: $n=200, 500, 1000$.
	\item[Factor 3] Number of covariates: $J=2$ and $J=5$. 
	\item[Factor 4] Estimators of $\theta_2 -\theta_1  $ and $\theta_3-\theta_1$ based on: (i) ANOVA estimator; (ii) an oracle ANHECOVA estimator under the ideal situation of no missing data, which cannot be computed in practice but is used as the benchmark;  (iii) ANHECOVA estimator with SI using the covariate-wise observed mean, denoted as \emph{SI (mean)}; (iv) ANHECOVA estimator with SI using  imputation values that are numerically optimized by \textsf{optim} in \textsf{R} to minimize the empirical variance in equation (S3) in the Supplementary Materials, denoted as \emph{SI (opt)}; (v) ANHECOVA estimator with MIM. We omit the ANHECOVA estimator with CWI using  imputation value $\widehat \bmc_{t}^\star$'s derived in Section \ref{sec:equi} because it is the same as the estimator in (v) by Lemma \ref{lemma: mim as cwi}.
\end{description} 
Throughout, we consider three treatment arms with allocation probabilities { $\pi_1=\pi_2=\pi_3=1/3$}.

 For Case 1, we generate $
\bmX = (X_{1},\dots, X_{5})^\top  \sim N((0.1, 0.2, 0.2, 0.3, 0.3)^\top, {\mathrm {diag}} \{ (2,2,1,2,1)\}),$ 
where ${\mathrm {diag}} \{\bm d\}$ denotes a diagonal matrix with the diagonal entries equal to $\bm d$. The potential outcomes follow $ Y^{(1)} =X_1^2-0.5 X_1+X_2+X_3^2-5 X_4+5 X_5+\epsilon_{1}, Y^{(2)} = 1.16-X_1+X_2^2-X_3+0.5 X_4+0.5 X_5+\epsilon_{2},Y^{(3)} = 3.85+X_1+X_2+X_3+0.5 X_4-X_5+\epsilon_3 $, 
where $\epsilon_1, \epsilon_2, \epsilon_3 \sim N(0,1)$. We then generate the missingness indicators $R_{j}, j=1,\dots, 5$ from a binomial distribution with probabilities $0.8, 0.7, 0.75, 0.65, 0.85$ respectively. 

For Case 2, $(\bmX, Y^{(1)}, Y^{(2)}, Y^{(3)}, Y)$ is generated the same way as in Case 1. We then generate the missingness indicators from a binomial distribution with probabilities $P(R_{j}=1\mid \bmX, Y^{(1)}, Y^{(2)}, Y^{(3)})=\exp(0.5 X_{j}-2)/\{1+\exp(0.5 X_{j}-2)\}$. 

For Case 3, we generate $\bmX$ from a multivariate normal distribution, with means and marginal variances the same as in Case 1, but $X_{j}$ and $X_{k}$, $j\neq k$ are correlated with correlation coefficient {$\rho=0.5$}. The potential outcomes follow $Y^{(1)} =X_1^2-0.5 X_1+X_2+X_3^2-0.5 X_{4}+X_{5}+\epsilon_1, Y^{(2)}  = 1.31-X_1+X_2^2-X_3+0.5 X_4+0.5X_5+\epsilon_2, Y^{(3)} = 4+X_1+X_2+X_3+0.5 X_4-X_5+\epsilon_3,$ where $\epsilon_1, \epsilon_2, \epsilon_3 \sim N(0,1)$. The missingness indicators are generated from a binomial distribution with probabilities {$ P(R_j=1\mid X_j, Y^{(1)}, Y^{(2)}, Y^{(3)}) =\exp\{0.2\cdot (Y^{(1)}+Y^{(2)}+Y^{(3)})-2 X_{j}-2\}/ \{1+\exp(0.2\cdot (Y^{(1)}+Y^{(2)}+Y^{(3)})-2 X_{j}-2)\}$ }.  {In this case, the condition in 
 Theorem \ref{lemma:equi} is violated.}

In all three cases, the observed outcome is $Y=I(\bmA=\bma_1) Y^{(1)}+I(\bmA=\bma_2) Y^{(2)}+I(\bmA=\bma_3) Y^{(3)}$, and the observed data are $(X_{i1}R_{i1},\dots X_{i5}R_{i5}, R_{i1},\dots, R_{i5}, A_i, Y_i), i=1,\dots, n$. The true treatment effects are \( \theta_2 - \theta_1 = 0 \) between arms 2 and 1, and \( \theta_3 - \theta_1 = 1 \) between arms 3 and 1. In addition, the true models for $Y^{(1)}$ and $Y^{(2)}$ have quadratic terms, while our working model is linear in all covariates, which is a misspecified model.

The results are in Table \ref{table:sim.null} and Table \ref{table:sim.alt}, which include (i) the Monte Carlo mean and standard deviation (SD) of each estimator, (ii) average of standard errors (SEs); { (iii) empirical type-I error rate when estimating $\theta_2- \theta_1$; (iv) empirical power when estimating $\theta_3- \theta_1$; and (v) coverage probability (CP) of 95\% confidence intervals from normal approximation when estimating $\theta_3- \theta_1$}. For situations with $J=2$, we only included $X_1, X_2$ in the model. After imputation, the ANOVA and ANHECOVA estimators are computed using the \textsf{R} package \textsf{RobinCar} \citep{robincar}. We use 5000 simulation runs {when estimating $\theta_2-\theta _1$ and 3000 simulation runs when estimating $\theta_3-\theta_1$}. 

\begin{table}[H]
\caption{Mean estimate, standard deviation (SD), average standard error (SE), and alpha (empirical type-I error rate) for $\theta_2 - \theta_1$ under simple randomization based on 5,000 simulations. The true value of $\theta_2 - \theta_1$ is 0. }
\label{table:sim.null}
\resizebox{\textwidth}{!}{
\begin{tabular}{cccccccccccccc}
\hline
\multicolumn{14}{c}{Case 1}                                                                                                        \\ \hline
                     &           & \multicolumn{4}{c}{$n=200$}     & \multicolumn{4}{c}{$n=500$}      & \multicolumn{4}{c}{$n=1000$}     \\ \cline{3-14} 
                     &           & mean  & SD    & SE    & alpha & mean   & SD    & SE    & alpha & mean   & SD    & SE    & alpha \\ \hline
$J=0$                  & ANOVA     & 0.004 & 1.253 & 1.254 & 0.054 & -0.020 & 0.798 & 0.792 & 0.057 & 0.001  & 0.561 & 0.560 & 0.055 \\ \hline
\multirow{4}{*}{$J=2$} & Oracle    & 0.004 & 1.238 & 1.215 & 0.061 & -0.023 & 0.784 & 0.773 & 0.056 & 0.003  & 0.550 & 0.548 & 0.052 \\
                     & SI (mean) & 0.005 & 1.243 & 1.221 & 0.057 & -0.022 & 0.789 & 0.777 & 0.057 & 0.004  & 0.552 & 0.550 & 0.051 \\
                     & SI (opt)  & 0.006 & 1.253 & 1.210 & 0.060 & -0.023 & 0.792 & 0.774 & 0.059 & 0.005  & 0.553 & 0.550 & 0.052 \\
                     & MIM       & 0.003 & 1.258 & 1.209 & 0.062 & -0.024 & 0.794 & 0.773 & 0.060 & 0.004  & 0.553 & 0.549 & 0.053 \\ \hline
\multirow{4}{*}{$J=5$} & Oracle    & 0.025 & 0.885 & 0.857 & 0.057 & -0.004 & 0.549 & 0.552 & 0.045 & {0.004} & 0.393 & 0.390 & 0.052 \\
                     & SI (mean) & 0.023 & 1.011 & 0.971 & 0.065 & -0.007 & 0.637 & 0.622 & 0.055 & 0.000  & 0.443 & 0.442 & 0.053 \\
                     & SI (opt)  & 0.022 & 1.037 & 0.939 & 0.079 & -0.009 & 0.642 & 0.615 & 0.060 & -0.001 & 0.445 & 0.440 & 0.056 \\
                     & MIM       & 0.025 & 1.034 & 0.953 & 0.073 & -0.008 & 0.642 & 0.617 & 0.059 & 0.001  & 0.445 & 0.440 & 0.058 \\ \hline
\multicolumn{14}{c}{Case 2}                                                                                                        \\ \hline
                     &           & \multicolumn{4}{c}{$n=200$}     & \multicolumn{4}{c}{$n=500$}      & \multicolumn{4}{c}{$n=1000$}     \\ \cline{3-14} 
                     &           & mean  & SD    & SE    & alpha & mean   & SD    & SE    & alpha & mean   & SD    & SE    & alpha \\ \hline
$J=0$                  & ANOVA     & \multicolumn{12}{c}{Same as Case 1}                     \\ \hline
\multirow{4}{*}{$J=2$} & Oracle    & \multicolumn{12}{c}{Same as Case 1}                                                             \\
                     & SI (mean) & 0.002 & 1.242 & 1.219 & 0.058 & -0.021 & 0.786 & 0.775 & 0.055 & 0.004  & 0.551 & 0.550 & 0.052 \\
                     & SI (opt)  & 0.015 & 1.248 & 1.206 & 0.063 & -0.017 & 0.788 & 0.772 & 0.057 & 0.006  & 0.551 & 0.548 & 0.051 \\
                     & MIM      & 0.005 & 1.257 & 1.205 & 0.066 & -0.022 & 0.789 & 0.771 & 0.059 & 0.004  & 0.551 & 0.548 & 0.052 \\ \hline
\multirow{4}{*}{$J=5$} & Oracle    & \multicolumn{12}{c}{Same as Case 1}                                                             \\
                     & SI (mean) & 0.021 & 0.971 & 0.936 & 0.061 & 0.000  & 0.608 & 0.600 & 0.056 & 0.003  & 0.430 & 0.427 & 0.053 \\
                     & SI (opt)  & 0.008 & 0.972 & 0.884 & 0.080 & -0.007 & 0.600 & 0.580 & 0.057 & -0.001 & 0.424 & 0.415 & 0.057 \\
                     & MIM      & 0.019 & 0.976 & 0.901 & 0.074 & -0.001 & 0.599 & 0.583 & 0.054 & 0.002  & 0.423 & 0.415 & 0.054 \\ \hline
\multicolumn{14}{c}{Case 3}                                                                                                        \\ \hline
                     &           & \multicolumn{4}{c}{$n=200$}     & \multicolumn{4}{c}{$n=500$}      & \multicolumn{4}{c}{$n=1000$}     \\ \cline{3-14} 
                     &           & mean  & SD    & SE    & alpha & mean   & SD    & SE    & alpha & mean   & SD    & SE    & alpha \\ \hline
$J=0$                  & ANOVA     & 0.008 & 0.674 & 0.672 & 0.048 & 0.001  & 0.421 & 0.427 & 0.043 & 0.004  & 0.303 & 0.303 & 0.050 \\ \hline
\multirow{4}{*}{$J=2$} & Oracle    & 0.019 & 0.653 & 0.628 & 0.056 & 0.000  & 0.402 & 0.405 & 0.048 & 0.007  & 0.292 & 0.288 & 0.054 \\
                     & SI (mean) & 0.025 & 0.604 & 0.590 & 0.059 & 0.002  & 0.381 & 0.378 & 0.052 & 0.006  & 0.271 & 0.269 & 0.053 \\
                     & SI (opt)  & 0.032 & 0.597 & 0.572 & 0.060 & 0.010  & 0.377 & 0.370 & 0.055 & 0.013  & 0.269 & 0.264 & 0.056 \\
                     & MIM      & 0.014 & 0.587 & 0.565 & 0.058 & -0.003 & 0.367 & 0.364 & 0.055 & 0.004  & 0.262 & 0.259 & 0.055 \\ \hline
\multirow{4}{*}{$J=5$} & Oracle    & 0.032 & 0.649 & 0.607 & 0.067 & 0.006  & 0.395 & 0.394 & 0.046 & 0.008  & 0.287 & 0.282 & 0.057 \\
                     & SI (mean) & 0.047 & 0.606 & 0.574 & 0.066 & 0.010  & 0.377 & 0.371 & 0.055 & 0.009  & 0.268 & 0.264 & 0.054 \\
                     & SI (opt)  & 0.052 & 0.585 & 0.525 & 0.084 & 0.021  & 0.362 & 0.347 & 0.061 & 0.018  & 0.258 & 0.250 & 0.061 \\
                     & MIM      & 0.033 & 0.573 & 0.522 & 0.080 & 0.005  & 0.350 & 0.340 & 0.056 & 0.006  & 0.249 & 0.244 & 0.052 \\ \hline
\end{tabular}}
\end{table}

\begin{table}[H]
\caption{Mean estimate, standard deviation (SD),  average standard error (SE), coverage probability (CP) of 95\% asymptotic confidence interval, and empirical power for $\theta_3 - \theta_1$ under simple randomization based on 3,000 simulations. The true value of $\theta_3 - \theta_1$ is 1.}
\label{table:sim.alt}
\resizebox{\textwidth}{!}{
\begin{tabular}{ccccccccccccccccc}
\hline
\multicolumn{17}{c}{Case 1}                                                                                                                              \\ \hline
                     &           & \multicolumn{5}{c}{$n=200$}             & \multicolumn{5}{c}{$n=500$}             & \multicolumn{5}{c}{$n=1000$}            \\ \cline{3-17} 
                     &           & mean  & SD    & SE    & CP    & Power & mean  & SD    & SE    & CP    & Power & mean  & SD    & SE    & CP    & Power \\ \hline
$J=0$                  & ANOVA     & 1.015 & 1.231 & 1.221 & 0.942 & 0.133 & 0.981 & 0.762 & 0.770 & 0.954 & 0.249 & 1.004 & 0.549 & 0.545 & 0.950 & 0.457 \\ \hline
\multirow{4}{*}{$J=2$} & Oracle    & 1.056 & 1.199  & 1.176& 0.944 & 0.155 & 0.994 & 0.750 &  0.747& 0.949 & 0.265 & 1.013 & 0.531 &  0.529& 0.949 & 0.484 \\
                     & SI (mean) & 1.058 & 1.212 & 1.185 & 0.940 & 0.155 & 0.996 & 0.755 & 0.752 & 0.946 & 0.262 & 1.013 & 0.537 & 0.533 & 0.950 & 0.482 \\
                     & SI (opt)  & 1.052 & 1.220 & 1.175 & 0.935 & 0.157 & 0.997 & 0.760  & 0.750 & 0.945 & 0.267 & 1.015 & 0.538 & 0.533 & 0.949 & 0.483 \\
                     & MIM      & 1.052 & 1.221 & 1.174 & 0.935 & 0.158 & 0.993 & 0.761 & 0.749 & 0.942 & 0.264 & 1.015 & 0.537 & 0.532 & 0.949 & 0.482 \\ \hline
\multirow{4}{*}{$J=5$} & Oracle    & 1.079 & 0.880 & 0.840 & 0.938 & 0.267 & 1.012 & 0.551 & 0.536 & 0.948 & 0.473 & 1.016 & 0.389 & 0.380 & 0.948 & 0.746 \\
                     & SI (mean) & 1.080 & 1.002 & 0.956 & 0.932 & 0.223 & 1.012 & 0.625 & 0.610 & 0.944 & 0.377 & 1.012 & 0.440 & 0.433 & 0.945 & 0.646 \\
                     & SI (opt)  & 1.064 & 1.026 & 0.925 & 0.916 & 0.242 & 1.011 & 0.630 & 0.603 & 0.935 & 0.387 & 1.011 & 0.441 & 0.430 & 0.945 & 0.648 \\
                     & MIM       & 1.078 & 1.023 & 0.940 & 0.921 & 0.237 & 1.011 & 0.628 & 0.605 & 0.938 & 0.390  & 1.013 & 0.441 & 0.430 & 0.943 & 0.646 \\ \hline
\multicolumn{17}{c}{Case 2}                                                                                                                              \\ \hline
                     &           & \multicolumn{5}{c}{$n=200$}             & \multicolumn{5}{c}{$n=500$}             & \multicolumn{5}{c}{$n=1000$}            \\ \cline{3-17}  
                     &           & mean  & SD    & SE    & CP    & Power & mean  & SD    & SE    & CP    & Power & mean  & SD    & SE    & CP    & Power \\ \hline
$J=0$                  & ANOVA     & \multicolumn{15}{c}{Same as Case 1}                     \\ \hline
\multirow{4}{*}{$J=2$} & Oracle    & \multicolumn{15}{c}{Same as Case 1}                                                                                   \\
                     & SI (mean) & 1.055 & 1.207 & 1.183 & 0.945 & 0.152 & 0.998 & 0.753 & 0.751 & 0.947 & 0.266 & 1.014 & 0.534 & 0.532 & 0.950 & 0.480 \\
                     & SI (opt)  & 1.056 & 1.208 & 1.171 & 0.940 & 0.157 & 0.996 & 0.756 & 0.748 & 0.943 & 0.269 & 1.014 & 0.535 & 0.531 & 0.949 & 0.489 \\
                     & MIM      & 1.058 & 1.215 & 1.169 & 0.938 & 0.162 & 0.996 & 0.758 & 0.746 & 0.943 & 0.273 & 1.014 & 0.535 & 0.530 & 0.949 & 0.485 \\ \hline
\multirow{4}{*}{$J=5$} & Oracle    & \multicolumn{15}{c}{Same as Case 1}                                                                                   \\
                     & SI (mean) & 1.070 & 0.960 & 0.532 & 0.935 & 0.224 & 1.018 & 0.602 & 0.587 & 0.946 & 0.410 & 1.016 & 0.423 & 0.417 & 0.948 & 0.671 \\
                     & SI (opt)  & 1.038 & 0.962 & 0.531 & 0.921 & 0.250  & 1.006 & 0.594 & 0.568 & 0.940 & 0.427 & 1.007 & 0.416 & 0.406 & 0.946 & 0.691 \\
                     & MIM       & 1.069 & 0.957 & 0.530 & 0.929 & 0.248 & 1.018 & 0.592 & 0.571 & 0.943 & 0.433 & 1.013 & 0.415 & 0.406 & 0.945 & 0.695 \\ \hline
\multicolumn{17}{c}{Case 3}                                                                                                                              \\ \hline
                     &           & \multicolumn{5}{c}{$n=200$}             & \multicolumn{5}{c}{$n=500$}             & \multicolumn{5}{c}{$n=1000$}            \\ \cline{3-17}  
                     &           & mean  & SD    & SE    & CP    & Power & mean  & SD    & SE    & CP    & Power & mean  & SD    & SE    & CP    & Power \\ \hline
$J=0$                  & ANOVA     & 1.012 & 0.690 & 0.669 & 0.937 & 0.348 & 1.000 & 0.426 & 0.423 & 0.951 & 0.645 & 1.006 & 0.304 & 0.299 & 0.946 & 0.910 \\ \hline
\multirow{4}{*}{$J=2$} & Oracle    & 1.066 & 0.567 & 0.540 & 0.928 & 0.522 & 1.015 & 0.353 & 0.346 & 0.944 & 0.810  & 1.014 & 0.249 & 0.246 & 0.938 & 0.972 \\
                     & SI (mean) & 1.021 & 0.580 & 0.556 & 0.937 & 0.466 & 0.995 & 0.363 & 0.354 & 0.937 & 0.785 & 1.004 & 0.256 & 0.251 & 0.943 & 0.967 \\
                     & SI (opt)  & 1.030 & 0.580 & 0.538 & 0.929 & 0.497 & 0.998 & 0.358 & 0.346 & 0.940 & 0.801 & 1.005 & 0.251 & 0.246 & 0.942 & 0.970 \\
                     & MIM       & 1.044 & 0.566 & 0.534 & 0.932 & 0.511 & 1.005 & 0.354 & 0.343 & 0.942 & 0.817 & 1.008 & 0.247 & 0.243 & 0.946 & 0.977 \\ \hline
\multirow{4}{*}{$J=5$} & Oracle    & 1.084 & 0.558 & 0.515 & 0.917 & 0.558 & 1.022 & 0.345 & 0.334 & 0.937 & 0.840 & 1.016 & 0.244 & 0.238 & 0.932 & 0.977 \\
                     & SI (mean) & 1.025 & 0.562 & 0.521 & 0.932 & 0.524 & 0.996 & 0.348 & 0.335 & 0.934 & 0.829 & 1.003 & 0.245 & 0.238 & 0.940 & 0.977 \\
                     & SI (opt)  & 1.027 & 0.562 & 0.491 & 0.911 & 0.561 & 0.997 & 0.345 & 0.323 & 0.926 & 0.841 & 1.003 & 0.240 & 0.231 & 0.940 & 0.983 \\
                     & MIM       & 1.051 & 0.550 & 0.491 & 0.914 & 0.582 & 1.007 & 0.338 & 0.320 & 0.928 & 0.857 & 1.008 & 0.235 & 0.228 & 0.939 & 0.987 \\ \hline
\end{tabular}}
\end{table}



The following is a summary of simulation results in Table \ref{table:sim.null} and Table \ref{table:sim.alt}.
\begin{enumerate}
	\item All estimators have negligible bias. {For large sample sizes ($n=1000$), the empirical type I error rates are close to the nominal level 0.05 in Table 1, and the coverage probabilities are close to the nominal level 0.95 in Table 2.} 
	\item Because all estimators apply ANHECOVA to adjust for covariates, using either SI or MIM generally results in a smaller standard deviation compared to using ANOVA. Additionally, increasing the number of covariates from 2 to 5 further reduces the standard deviation. {This pattern holds for both tables.} In addition, Table 2 shows that adjusting for covariates results in higher power, and adjusting for more covariates further increases power.	
	\item {In Case 1, which is MCAR, the condition in Theorem  \ref{lemma:equi} holds, and our Corollary
	\ref{corollary:MCAR} implies that SI (mean), SI (opt), and MIM  have the same asymptotic efficiency.} Indeed, for large sample sizes ($n=1000$), we observe that these three methods have similar standard deviation {in both tables}, which is consistent with our asymptotic theory in Theorem  \ref{lemma:equi}  and Corollary \ref{corollary:MCAR}. 
	\item For large sample sizes ($n=1000$), MIM has the smallest standard deviations and standard errors across all scenarios in both tables, which is consistent with our asymptotic theory in Theorem \ref{theorem:optimality}. {Specifically, in Case 3, MIM clearly has smaller  standard deviations compared to the other methods in both tables, and larger power compared to the other methods in Table 2.}
	\item Under Case 1 and 2, the oracle estimator has smaller standard derivations compared to the SI and MIM methods. Interestingly, in Case 3, the oracle estimator can have larger standard deviations than both the SI and MIM estimators. This occurs because the SI and MIM methods also utilize the information from the missing indicators, which are associated with the outcome.
	\item 
	When the sample size is small ($n=200$) and the number of covariates is relatively large ($J=5$), the standard errors of SI  (opt) and MIM underestimate their standard deviations so that the empirical type I error rates are around 7.4\%-8.4\% in Table 1 and coverage probabilities are around 91\%-93\%. In comparison, the SI (mean) has better type I error control and higher coverage probabilities. This is because when $J=5$, MIM actually adjusts for 10 regressors (5 for covariates, 5 for missingness indicators) while SI (mean) only adjusts for 5 regressors. Moreover, SI (mean) with $J=5$ has similar or smaller standard deviation than MIM with $J=2$, suggesting that if the number of regressors is limited by a fixed sample size, adjusting for two times the covariates with SI may lead to higher efficiency compared to MIM. 
\end{enumerate}

%% file: sections_arxiv/06_realdata.tex

\section{Real Data Application}\label{sec: real data}
We apply the covariate adjustment methods to the Childhood Adenotonsillectomy Trial (CHAT) \citep{marcuschat, zhangchat}. As reviewed in Section 1.2, CHAT is a two-arm randomized controlled trial, which was designed to evaluate the efficacy of early adenotonsillectomy compared to watchful waiting with supportive care in children with the obstructive sleep apnea syndrome. The primary outcome of CHAT is the change in the attention and executive-function score, {which we use as the outcome in our analysis.} We focus on  $n=397$ subjects with non-missing outcomes. Of these, 203 subjects are in the watchful waiting with supportive care group, and 194 are in the adenotonsillectomy group. We consider baseline BMI, gender, tonsil size, PSQ score, and OSA-18 score as covariates for adjustment. The missing proportion of covariates and the correlation between each covariate and the outcome are shown in Table \ref{table:desc}. 

\begin{table}[H]
\caption{Missing rates of covariates, and marginal correlations between each covariate (and its missingness indicator) and the outcome in CHAT}
\centering
\label{table:desc}
\begin{tabular}{@{} cccc @{}}
\toprule
Covariates & Missing rate & $\mathrm{Cor}(X, Y)$& $\mathrm{Cor}(R, Y)$ \\ \midrule
PSQ  & 0.755\% & -0.024 & -0.198 \\ \midrule
{OSA-18}  & 0.504\% & -0.033 & 0.000 \\ \midrule
BMI & 3.274\% & 0.010 & 0.058 \\ \midrule
Gender & 0.000\% & 0.056 & 0.000 \\ \midrule
Tonsil size & 0.755\% & 0.081 & -0.065 \\ \bottomrule
\end{tabular}

\end{table}

We compare two covariate adjustment sets - one with tonsil size and gender (the two variables with the highest correlations with the outcome), and the other with tonsil size, gender, BMI, OSA score, and PSQ score. We fitted ANOVA, SI (mean), SI (opt), and MIM, all as defined and evaluated in simulations in Section \ref{sec: sim}. Additionally, we conducted complete case analysis using ANOVA and AHNECOVA, which exclude all observations with missing values. Complete case analysis is generally not recommended as it contradicts the  intention-to-treat principle and is both biased and inefficient \citep{whiteandthompson}. The results are summarized in Table \ref{table:real}. 

\begin{table}[H]
\centering
\caption{Real data results for CHAT}
\label{table:real}
\resizebox{!}{!}{
\begin{tabular}{ccccc}
\hline
\multicolumn{3}{c}{}  & Estimate   & SE    \\ \hline
\multirow{6}{*}{\begin{tabular}[c]{@{}c@{}}$J=2$\\ Tonsil size + gender\end{tabular}} & \multirow{4}{*}{\begin{tabular}[c]{@{}c@{}}With missing\\ $n=397$\end{tabular}}  & ANOVA  & 1.914 & 1.375 \\ &   & SI (mean) & 1.741 & 1.370 \\  &  & SI (opt)     & 1.662 & 1.367 \\    &    & MIM   & 1.705 & 1.367 \\ \cline{2-5}     & \multirow{2}{*}{\begin{tabular}[c]{@{}c@{}}Complete Case\\ $n=394$\end{tabular}} & ANOVA  & 1.707 & 1.373 \\  &     & ANHECOVA   & 1.546 & 1.368 \\ \hline
\multirow{6}{*}{$J=5$}   & \multirow{4}{*}{\begin{tabular}[c]{@{}c@{}}With missing\\ $n=397$\end{tabular}}  & ANOVA             & 1.914 & 1.375 \\    &     & SI (mean) & 1.745 & 1.370 \\    &     & SI (opt)     & 1.475 & 1.340 \\    &     & MIM    & 1.530 & 1.342 \\ \cline{2-5}     & \multirow{2}{*}{\begin{tabular}[c]{@{}c@{}}Complete Case\\ $n=377$\end{tabular}} & ANOVA  & 0.989 & 1.367 \\    &  & ANHECOVA   & 0.863 & 1.362 \\ \hline
\end{tabular}}
\end{table}
From Table \ref{table:real}, we can see that ANOVA, SI (mean), SI (opt), and MIM have similar effect sizes, regardless of the number of covariates adjusted. Among these methods, SI (opt) and MIM have the smallest standard errors, 
and including more covariates further reduces standard errors. In contrast, adjusting more covariates using SI (mean) has little impact on the standard errors. 

With two covariates exhibiting very low missing rates (shown in Table 3), the two complete case analysis methods exclude only three individuals and still provide point estimates similar to  those from the other correct analytical approaches. However, when adjusting for five covariates, the two complete case analysis methods give point estimates that are very different from those of the other correct analytical methods due to bias. Moreover, the  two complete case analysis methods have larger standard errors than SI(opt) and MIM, both for $J=2$ and $J=5$.


%% file: sections_arxiv/07_discussion.tex

\section{Discussion}\label{sec: discussion}
We proposed and investigated a cross-world imputation framework, using which we compare the efficiency between the single imputation and MIM methods. We showed that MIM implicitly searches for the optimal CWI values and thus achieves optimal efficiency in the asymptotic sense. We also derived conditions under which the single imputation method, by searching for the optimal single imputation values, can achieve the same asymptotic efficiency as the MIM. In particular, we showed that SI by covariate-wise observed means can achieve the same asymptotic efficiency as the MIM under missing completely at random.  


Based on our results, below are our general recommendations in choosing strategies to adjust for incomplete baseline covariates: 
\begin{enumerate}
\item If the condition in Theorem 3 holds, then asymptotically MIM and optimal SI achieve the same efficiency gain. This condition holds under the following two scenarios:
(a) The absence of treatment-by-covariate and treatment-by-missingness interactions in the outcome generating process;
(b) MCAR, i.e., $\bmR \perp (\bmA, \bmX, Y^{(t)} )$ for all $t$. {Furthermore, the condition in Theorem 3 can be empirically verified using historic data and the explicit forms of $\lim_{n\to \infty}\tilde \gamma_{tj} (\bm{0})$ and $ \lim_{n\to \infty}\tilde \beta_{tj} (\bm{0})$ given in the Supplementary Materials.

}
\item Under MCAR,  the optimal imputation values for SI is the covariate-wise observed means (Corollary 1).  Hence, under MCAR, SI  by covariate-wise observed means is recommended for its optimal asymptotic efficiency and better finite-sample performance than MIM. 
\item When MCAR does not hold,  the common practice of SI  by covariate-wise observed means is not necessarily an efficient strategy. For example, consider a scenario with a single covariate, two treatments, and estimation of the ATE. If $\cov\big\{ RX + (1- R) E(X\mid R=1), \pi_1Y^{(2)}+\pi_2Y^{(1)}\big\} = 0$, SI  by the observed mean $E(X\mid R=1)$ 
is the least efficient choice, as shown in the Supplement S3.
\item If the number of regressors is limited by a fixed sample size, adjusting for two times the covariates with SI may lead to higher efficiency compared to MIM, while maintaining the same number of regressors.
\item  Lastly, it is crucial to pre-specify the method for handling missing covariate values, based on prior knowledge of the aforementioned conditions or historical data, to ensure the validity of the inferences.
\end{enumerate}

%% file: sections_arxiv/08_ac.tex

We would like to thank Peng Ding for helpful comments and suggestions. We also thank the Associate Editor and three anonymous referees for helpful comments and suggestions. The Childhood Adenotonsillectomy Trial (CHAT) was supported by the National Institutes of Health (HL083075, HL083129, UL1-RR-024134, UL1 RR024989). The National Sleep Research Resource was supported by the National Heart, Lung, and Blood Institute (R24 HL114473, 75N92019R002). This work was supported by National Institute of Allergy and Infectious Diseases [NIAID 5 UM1 AI068617].

%% file: sections_arxiv/10_dataavailability.tex
\section*{Data Availability}
The Childhood Adenotonsillectomy Trial data that support the findings in this paper is publicly available at the National Sleep Research Resource database at \url{https://sleepdata.org/datasets/chat/pages}

%% file: sections_arxiv/09_supp.tex
\setcounter{equation}{0}
\setcounter{table}{0}
\setcounter{lemma}{0}
\setcounter{section}{0}
\renewcommand{\theequation}{S\arabic{equation}}
\renewcommand{\thecorollary}{S\arabic{corollary}}
\renewcommand{\thetable}{S\arabic{table}}
\renewcommand{\thefigure}{S\arabic{figure}}
\renewcommand{\thesection}{S\arabic{section}}
\renewcommand{\thelemma}{S\arabic{lemma}}

\section{Additional theoretical results}
\subsection{Asymptotic variance in Theorem 1}
As stated in  Theorem 1, the asymptotic distribution of $\widehat{\bmtheta}(\widehat{\bb}_1, \dots, \widehat \bb_k; \widehat{\bmc}_1, \dots, \widehat{\bmc}_k) $ is 
$$
\sqrt{n}\Big(  \widehat{\bmtheta}(\widehat{\bb}_1, \dots, \widehat \bb_k; \widehat{\bmc}_1, \dots, \widehat{\bmc}_k)-\bmtheta\Big)\xrightarrow{d}  N\Big(0, V(\bm{B}; \bmc_1, \dots, \bmc_k )\Big),
$$
where
	\begin{align*}
	V(\bm{B}; \bmc_1, \dots, \bmc_k ) & ={\mathrm{diag}}\Big[\pi_t^{-1}\var\{Y^{(t)}-\bb_t^\top \impx (\bmc_t)\}\Big]-\bm{B} \bm{\Sigma}_X(\bmc_1,\dots,\bmc_k) \bm{B}^\top \\
	&\quad +\bm{\mathscr{B}} \bm{D}_X(\bmc_1,\dots,\bmc_k) \bm{B}^\top + \bm{B} \bm{D}_X(\bmc_1,\dots,\bmc_k) \mathscr{\bm B}^\top ,
	\end{align*}
	${\mathrm{diag}}(d_t)$ denotes a $k\times k$ diagonal matrix with the $t$-th diagonal element $d_t$, {\small
	\begin{align*}
	& \bm{\Sigma}_X(\bmc_1,\dots,\bmc_k) =\begin{pmatrix}
	\var( \impx (\bmc_1)) & \cov(\impx(\bmc_1),\impx(\bmc_2)) & \cdots & \cov(\impx(\bmc_1),\impx(\bmc_k))\\
	\cov(\impx(\bmc_2),\impx(\bmc_1))& \var(\impx(\bmc_2)) & \cdots & \cov(\impx(\bmc_2),\impx(\bmc_k))\\
	\vdots & \vdots & \ddots & \vdots\\
	\cov(\impx(\bmc_k),\impx(\bmc_1)) & \cov(\impx(\bmc_k),\impx(\bmc_2))& \cdots & \var(\impx(\bmc_k))
	\end{pmatrix},\\
	& \bm{D}_X(\bmc_1,\dots,\bmc_k) =\begin{pmatrix}
	\var(\impx(\bmc_1)) & 0 & \cdots & 0\\
	0& \var(\impx(\bmc_2)) & \cdots & 0\\
	\vdots & \vdots & \ddots & \vdots\\
	0 & 0& \cdots & \var(\impx(\bmc_k))
	\end{pmatrix},\\
	& \mathscr{B}  =\begin{pmatrix}
	\bbeta_1(\bmc_1)^\top  & \bbeta_1(\bmc_2)^\top & \cdots & \bbeta_1(\bmc_k)^\top\\
	\bbeta_2(\bmc_1)^\top & \bbeta_2(\bmc_2)^\top & \cdots & \bbeta_2(\bmc_k)^\top\\
	\vdots & \vdots & \ddots & \vdots\\
	\bbeta_k(\bmc_1)^\top & \bbeta_k(\bmc_2)^\top & \cdots & \bbeta_k(\bmc_k)^\top
	\end{pmatrix}, \quad B=\left(
	\begin{array}{cccc}
	\bb_1^{\top}& 0 &\cdots & 0 \\
	0& \bb_2^{\top} & \vdots&0\\
	\vdots & \vdots & \ddots & \vdots\\
	0& 0& 0 & \bb_k^{\top}
	\end{array}\right),
	\end{align*}}
	and $\bmbeta_t(\bmc_s)= \var (\impx(\bmc_s))^{-1}  \cov (\impx(\bmc_s), Y^{(t)} )$.

Linear contrasts of outcome means in two arms are often of interest in many clinical trials. The following corollary provides the asymptotic property of linear contrasts using ANHECOVA.

\begin{corollary} \label{theo: asymp.cwi}
	Under the same conditions as in Theorem \ref{theo:asymp}, 
	\begin{align}
	&\sqrt{n} \big(       \widehat\theta^{\mathrm{cwi}}_{_{\mathrm{ANHC},t}} (\widehat \bmc_t) -         \widehat\theta^{\mathrm{cwi}}_{_{\mathrm{ANHC},s}} (\widehat \bmc_s)  - (\theta_t - \theta_s) \big)  \xrightarrow{d} N\big( 0, \sigma_{{{ts}}}^2(\bmc_t,\bmc_s) \big),  \label{eq: anhc}
	\end{align}
	for $t, s = 1,\dots, k$, where 
	\begin{align*}
	\sigma_{{ts}}^2(\bmc_t,\bmc_s) &= \frac{\var \{Y^{(t)}  - \bbeta_t (\bmc_t)^\top \bimpx (\bmc_t)\}}{\pi_t }  + \frac{\var \{Y^{(s)}  - \bbeta_s (\bmc_s)^\top\bimpx(\bmc_s)\}}{\pi_s} \\
	& \quad + \bbeta_t(\bmc_t)^\top \var ( \bimpx(\bmc_t))  \bbeta_t(\bmc_t) + \bbeta_s(\bmc_s)^\top \var ( \bimpx(\bmc_s))  \bbeta_s(\bmc_s) \\
	& \quad + 2\bbeta_t(\bmc_t) ^\top \cov(\bimpx(\bmc_t), \bimpx(\bmc_s))\bbeta_s(\bmc_s) \\
	& \quad -2\bbeta_s(\bmc_t)^\top\var(\bimpx(\bmc_t))\bbeta_t(\bmc_t) -2\bbeta_s(\bmc_s)^\top\var({\bimpx(\bmc_s)})\bbeta_t(\bmc_s) .
	\end{align*}
\end{corollary}

Let $ S_{t}^2(\widehat \bmc_t)$ be the sample variance of $ Y_i - \widehat\bbeta_t(\widehat \bmc_t)^\top  \bimpx_i(\widehat \bmc_t)  $'s in the treatment arm $ t $. Let $ \widehat{ \bm{\Sigma} }(\widehat \bmc_t) $ be the sample variance of $ \impx_i(\widehat \bmc_t) $'s,  and $\widehat{\bm \Sigma}_{ts}(\widehat \bmc_t,\widehat \bmc_s)$ be the sample covariance of $ \impx_i(\widehat \bmc_t) $'s and $ \impx_i(\widehat \bmc_s) $'s.
For model-free inference on the linear contrast, the following estimators are consistent for $\sigma_{{ts}}^2(\bmc_t,\bmc_s) $,
\begin{align*}
\widehat{\sigma}_{{ts}}^2(\widehat \bmc_t,\widehat \bmc_s) &= \frac{S_{t}^2  (\widehat \bmc_t)}{\pi_t} +   \frac{S_{s}^2  (\widehat \bmc_s)}{\pi_s} + \widehat{\bbeta}_t(\widehat \bmc_t)^\top \widehat{\bm\Sigma}(\widehat \bmc_t) \widehat{\bbeta}_t(\widehat \bmc_t) + \widehat{\bbeta}_s(\widehat \bmc_s)^\top \widehat{\bm\Sigma}(\widehat \bmc_s) \widehat{\bbeta}_s(\widehat \bmc_s) \\
& \quad + 2\widehat{\bbeta}_t(\widehat \bmc_t) ^\top \widehat{\bm\Sigma}_{ts}(\widehat \bmc_t,\widehat \bmc_s) \widehat{\bbeta}_s(\widehat \bmc_s)-2\widehat{\bbeta}_s(\widehat\bmc_t)^\top\widehat{\bm\Sigma}(\widehat \bmc_t)\widehat{\bbeta}_t(\widehat\bmc_t)-2\widehat{\bbeta}_s(\widehat\bmc_s)^\top\widehat{\bm\Sigma}(\widehat \bmc_s)\widehat{\bbeta}_t(\widehat\bmc_s).
\end{align*}

As CWI includes SI as special cases, Theorem \ref{theo:asymp} and Corollary S1 directly imply the asymptotic properties of the SI method. In particular, for SI with $\widehat\bmc$, where $\widehat \bmc=\bmc +o_p(1)$, the asymptotic variance of $\sqrt{n} \big(       \widehat\theta^{\mathrm{si}}_{_{\mathrm{ANHC}, t}} (\widehat \bmc) -     \widehat\theta^{\mathrm{si}}_{_{\mathrm{ANHC}, s}} (\widehat \bmc)  - (\theta_t - \theta_s) \big) $  is 
\begin{align}
\sigma_{{ts}}^2(\bmc) = &\frac{\var \{Y^{(t)}  - \bbeta_t (\bmc)^\top \bimpx (\bmc)\}}{\pi_t }  + \frac{\var \{Y^{(s)}  - \bbeta_s (\bmc)^\top\bimpx(\bmc)\}}{\pi_s} \nonumber\\
&\quad + \{\bbeta_t(\bmc) - \bbeta_s(\bmc) \}^\top  \var ( \bimpx(\bmc)) \{\bbeta_t(\bmc) - \bbeta_s(\bmc) \},\label{eq: si anhc asymp} 
\end{align} 
which can be consistently estimated by 
\begin{align}
\widehat 	\sigma_{{ts}}^2(\widehat \bmc) =\frac{S_{t}^2  (\widehat \bmc)}{\pi_t} +   \frac{S_{s}^2  (\widehat \bmc)}{\pi_s} + \{\widehat \bbeta_t(\widehat \bmc) - \widehat \bbeta_s(\widehat \bmc) \}^\top \widehat{\bm\Sigma}(\widehat \bmc)\{\widehat \bbeta_t(\widehat \bmc) - \widehat \bbeta_s(\widehat \bmc) \}. \label{eq: si anhc} 
\end{align}

\subsection{Optimal single imputation value under a special case}\label{supp:opt.imp}
We consider the special case with a single covariate,
two treatments, and estimating the average treatment effect $\theta_2- \theta_1$ in more detail. In this case,   $  - \lim_{n\to \infty}\widetilde \gamma_{t} (0)/ \lim_{n\to \infty}\widetilde \beta_{t} (0)
$  in Theorem \ref{lemma:equi} can be simplified to 
\[
- \frac{\var (XR) \cov (R, Y^{(t)}) - \cov (XR, R) \cov (XR, Y^{(t)})}{\var (R) \cov (XR, Y^{(t)}) - \cov (XR, R) \cov (R, Y^{(t)})} ,
\]
which does not depend on $t$ when $\cov (R, Y^{(t)}) /\cov(RX,Y^{(t)}) $ does not depend on $t$. This implies that when $R$ is independent of $(Y^{(1)}, Y^{(2)})$, SI can achieve optimality among CWI and the optimal imputation value is $\cov (XR, R)/\var (R) = E(X| R=1)$. In other words,
in this situation, imputing using the average observed covariate values leads to the optimal ANHECOVA estimator. In general, the optimal imputation value for SI can be obtained from minimizing the asymptotic variance of $\sqrt{n} \left(       \widehat\theta^{\mathrm{si}}_{_{\mathrm{ANHC}, 2}} (\widehat c) -     \widehat\theta^{\mathrm{si}}_{_{\mathrm{ANHC}, 1}} (\widehat c)  - (\theta_2 - \theta_1) \right) $ given by Corollary S1, which can be simplified to
\begin{align*}
\sigma_{{ts}}^2(c)=\frac{\pi_1 \var(Y^{(2)}) + \pi_1\var(Y^{(1)})  }{\pi_1\pi_2} - \frac{ \{\cov (RX, \pi_1Y^{(2)}+ \pi_2Y^{(1)} ) +\cov (1-R, \pi_1Y^{(2)}+ \pi_2Y^{(1)} ) c \}^2  }{\pi_1 \pi_2 \var (RX + (1-R) c) } .
\end{align*}
Here, we can see that $ 	\sigma_{{ts}}^2(c)$  is not a convex function of $c$. Again, it is easy to see that when $R$ is independent of $(Y^{(1)}, Y^{(2)})$, the numerator in the second term is free of $c$, and the denominator in the second term is minimized at $c= E(X\mid R=1)$. Hence, $	\sigma_{{ts}}^2(c)$ is minimized at  $c= E(X\mid R=1)$. When $R$ is correlated with $(Y^{(1)}, Y^{(2)})$, we define 
$\tau_{RX}=\cov(RX,\pi_1Y^{(2)}+\pi_2Y^{(1)})$ and $\tau_{R}=\cov(1-R,\pi_1Y^{(2)}+\pi_2Y^{(1)})$. As shown in the Supplementary Materials, when
$\tau_{RX}+\tau_R\cdot \E(X|R=1)\neq 0$, the optimal imputation value for SI is 
$$
c =\frac{\tau_R  \{\E(R)\var(X|R=1)+\var(R)\E(X|R=1)^2\}+\tau_{RX} \var(R)\E(X|R=1)}{\var(R)\E(X|R=1) \tau_R+\var(R) \tau_{RX}};
$$ 
when $\tau_{RX}+\tau_R  \E(X|R=1)= 0$, imputing using a pre-specified large number leads to near optimal efficiency. The argument based on optimizing the asymptotic variance can be easily generalized to the case when we have multiple mutually independent covariates, but becomes complex in the general case because the  asymptotic variance is not a convex function of $\bmc$. Nonetheless,
the optimal imputation value that minimizes $	\sigma_{{ts}}^2(\bmc)$ can be obtained using numerical optimization, e.g., using the \textsf{optim} package in \textsf{R}.

\subsection{Three Lemmas}

Next, we state three lemmas that will be useful in our proof.

Lemma \ref{lemma: beta} derives the probability limits of $\widehat \bbeta_t (\widehat \bmc_t)$.
\begin{lemma} \label{lemma: beta}
Under the same conditions in Theorem \ref{theo:asymp}, $\widehat \bbeta_t (\widehat \bmc_t) = \bbeta_t (\bmc_t) + o_p(1)$ and $\widehat\bbeta (\widehat\bmc) = \bbeta (\bmc) + o_p(1) $, where $\bbeta(\bmc) = \sum_{t=1}^k \pi_t \bbeta_t (\bmc)$. 
\end{lemma}

When there are two treatment arms, Lemma \ref{lemma: treatment-specific} shows that imputation by overall observed mean returns a more efficient ANHECOVA estimator than imputation by treatment-specific observed mean. In this case, 
imputation by treatment-specific values uses $ \widehat \bmc_t $ for subjects under treatment $ t $, and  $ \widehat \bmc_t= \bmc_t+o_p(1)$  for $t=1,2$. The resulting ANHECOVA estimator is 
\begin{align*}
	& {\widehat\theta_{{}_{\mathrm{ANHC,2}}}^{\mathrm {ti}}(\widehat \bmc_1, \widehat \bmc_2)-\widehat\theta_{{}_{\mathrm{ANHC,1}}}^{\mathrm {ti}}(\widehat \bmc_1, \widehat \bmc_2)} \\
	& =  \overline Y_2  -   \overline Y_ 1   - \widehat\bbeta_2(\widehat \bmc_2)^\top  ( \overline{\bmX}_2^{\mathrm {imp}} (\widehat \bmc_2)  - \overline{\bmX}^{\mathrm {imp}} (\widehat \bmc_1, \widehat \bmc_2) ) + \widehat\bbeta_1(\widehat \bmc_1)^\top  ( \overline{\bmX}_1^{\mathrm {imp}} (\widehat \bmc_1)  - \overline{\bmX}^{\mathrm {imp}} (\widehat \bmc_1, \widehat \bmc_2) ),
\end{align*}
where $\widehat\bbeta_t(\widehat \bmc_t), t=1,2$ is defined before Theorem \ref{theo:asymp}, and $\overline{\bmX}^{\mathrm {imp}} (\widehat \bmc_1, \widehat \bmc_2)  = \frac{n_2}{n}  \overline{\bmX}_2^{\mathrm {imp}} (\widehat \bmc_2)  + \frac{n_1}{n}  \overline{\bmX}_1^{\mathrm {imp}} (\widehat \bmc_1)$. 
When $\widehat \bmc_1 $ and $ \widehat \bmc_2$ are equal, i.e., when $\widehat \bmc_1 = \widehat \bmc_2 =  \widehat \bmc$, we have {$\widehat\theta_{{}_{\mathrm{ANHC,2}}}^{\mathrm {ti}}(\widehat \bmc_1, \widehat \bmc_2)-\widehat\theta_{{}_{\mathrm{ANHC,1}}}^{\mathrm {ti}}(\widehat \bmc_1, \widehat \bmc_2)= \widehat\theta_{{}_{\mathrm{ANHC,2}}}^{\mathrm {si}} (\widehat \bmc)-\widehat\theta_{{}_{\mathrm{ANHC,2}}}^{\mathrm {si}} (\widehat \bmc)$. }

\begin{lemma}\label{lemma: treatment-specific}
Under the same conditions in Theorem \ref{theo:asymp}, let $\widehat \bmc_1$ and $\widehat\bmc_2$ be the treatment-specific observed covariate means, and $\widehat\bmc$ be the overall observed covariate means, then the asymptotic variance of  {$${\widehat\theta_{{}_{\mathrm{ANHC,2}}}^{\mathrm {ti}}(\widehat \bmc, \widehat \bmc)-\widehat\theta_{{}_{\mathrm{ANHC,1}}}^{\mathrm {ti}}(\widehat \bmc, \widehat \bmc)}= \widehat\theta_{{}_{\mathrm{ANHC,2}}}^{\mathrm {si}} (\widehat \bmc) - \widehat\theta_{{}_{\mathrm{ANHC,1}}}^{\mathrm {si}} (\widehat \bmc)$$ is no larger than the asymptotic variance of  	${\widehat\theta_{{}_{\mathrm{ANHC,2}}}^{\mathrm {ti}}(\widehat \bmc_1, \widehat \bmc_2)-\widehat\theta_{{}_{\mathrm{ANHC,1}}}^{\mathrm {ti}}(\widehat \bmc_1, \widehat \bmc_2)}$.}
\end{lemma}

For ease of reading,  we re-state a useful Lemma in \cite{zhao2022adjust}. This lemma guarantees the invariance of least squares to non-degenerate linear transformation of the design matrix.
\begin{lemma}[\cite{zhao2022adjust}, Lemma S4]\label{lemmas:ding}
Consider an $n\times 1$ vector $Y$ and two $n\times Q$ matrices, $\bm X_1$ and $\bm X_2$, that satisfy $\bm X_2= \bm X_1 \bm \Gamma$ for some invertible $Q\times Q$ matrix  $\bm \Gamma$. The least squares fits 
\[
Y= \bmX_1\widehat\bbeta_1 + \widehat{\bm\epsilon}_1, \quad Y= \bmX_2\widehat\bbeta_2 + \widehat{\bm \epsilon}_2
\]
satisfy $\widehat\bbeta_1 = \bm\Gamma \widehat\bbeta_2$ and $\widehat{\bm\epsilon}_1 = \widehat{\bm\epsilon}_2$.
\end{lemma}

\section{Technical Proofs}

\subsection{Proof of Lemma \ref{lemma: beta}} 
 
Recall the definition that 
 \begin{align*}
	\widehat\bbeta_t (\widehat \bmc_t) &=  \bigg\{ \sum_{ i : A_i =a_t} (\bimpx_i(\widehat \bmc_t) - \overline{\bmX}_t^{\mathrm {imp}} (\widehat \bmc_t) )(\bimpx_i(\widehat \bmc_t) - \overline{\bmX}_t^{\mathrm {imp}} (\widehat \bmc_t) )^\top \bigg\}^{-1}   \\
	&\quad \quad \quad\quad\quad \sum_{ i : A_i =a_t}  (\bimpx_i(\widehat \bmc_t) - \overline{\bmX}_t^{\mathrm {imp}} (\widehat \bmc_t) )Y_i  .
\end{align*}
Since $\widehat{\bmc}_t=\bmc_t+o_p(1)$, we have 
\begin{align*}
\overline{\bmX}_t^{\mathrm {imp}} (\widehat \bmc_t) &= \frac{1}{n_t}\sum_{ i : A_i =a_t}(\bmX_i \circ \bmR_i+\widehat{\bmc}_t \circ (\bm 1-\bmR_i))\\
&=\frac{1}{n_t}\sum_{ i : A_i =a_t}(\bmX_i \circ \bmR_i+\bmc_t \circ (\bm 1-\bmR_i))+ (\widehat \bmc_t - \bmc_t) \circ \frac{1}{n_t} \sum_{ i : A_i =a_t}  (\bm 1-\bmR_i)   \\
&= \frac{1}{n_t}\sum_{ i : \bmA_i =\bma_t}\bmX_i^{\mathrm{imp}}(\bmc_t)+o_p(1)\\
&= E(\bmX^{\mathrm{imp}}(\bmc_t)\mid \bmA=\bma_t)+o_p(1)\\
&= E (\bimpx (\bmc_t)) +o_p(1),
\end{align*}
where the third equality is because  $\frac{1}{n_t} \sum_{ i : A_i =a_t}  (1-\bmR_i) = E(1- R_i \mid \bmA =\bma_t) + o_p(1) $. Similarly, we have $ \overline Y_t= E(Y^{(t)}) + o_p(1),$ and $\frac{1}{n_t} \sum_{ i : A_i =a_t}  \bimpx_i(\widehat \bmc_t)  Y_i  = E ( \bimpx (\bmc_t) Y^{(t)}  ) + o_p(1)$.  Then, we have  
\begin{align*}
	\frac{1}{n_t} \sum_{ i : A_i =a_t}  (\bimpx_i(\widehat \bmc_t) - \overline{\bmX}_t^{\mathrm {imp}} (\widehat \bmc_t) )Y_i  & =  \frac{1}{n_t} \sum_{ i : A_i =t}  \bimpx_i(\widehat \bmc_t)  Y_i - \overline{\bmX}_t^{\mathrm {imp}} (\widehat \bmc_t) \overline Y_t \\
	&= \cov ( \bimpx (\bmc_t), Y^{(t)}) + o_p(1).
\end{align*}
Similarly, we can show that  
\begin{align*}
	\frac{1}{n_t}\sum_{ i : A_i =t} \{\bimpx_i(\widehat \bmc_t) - \overline{\bmX}_t^{\mathrm {imp}} (\widehat \bmc_t) \}\{\bimpx_i(\widehat \bmc_t) - \overline{\bmX}_t^{\mathrm {imp}} (\widehat \bmc_t) \}^\top = \var (\bmX^{\mathrm{imp} }(\bmc_t))+o_p(1). 
\end{align*}
Combining these arguments, we have that $ \widehat\bbeta_t(\widehat \bmc_t ) = \bbeta_t(\bmc_t) +o_p(1)$. Similarly, we can prove that $ \widehat\bbeta(\widehat \bmc) = \bbeta(\bmc) +o_p(1)$.

\subsection{Proof of Lemma \ref{lemma: treatment-specific}} 
\label{supp sec: proof of treatment-arm specific imputation}

Recall that the ANHECOVA estimator can be written as
\begin{align*}
	& {\widehat\theta_{{}_{\mathrm{ANHC,2}}}^{\mathrm {ti}}(\widehat \bmc_1, \widehat \bmc_2)-\widehat\theta_{{}_{\mathrm{ANHC,1}}}^{\mathrm {ti}}(\widehat \bmc_1, \widehat \bmc_2)} \\
	& =  \overline Y_2  -   \overline Y_ 1   - \widehat\bbeta_2(\widehat \bmc_2)^\top  ( \overline{\bmX}_2^{\mathrm {imp}} (\widehat \bmc_2)  - \overline{\bmX}^{\mathrm {imp}} (\widehat \bmc_1, \widehat \bmc_2) ) + \widehat\bbeta_1(\widehat \bmc_1)^\top  ( \overline{\bmX}_1^{\mathrm {imp}} (\widehat \bmc_1)  - \overline{\bmX}^{\mathrm {imp}} (\widehat \bmc_1, \widehat \bmc_2) ) \\
	&= \overline Y_2 - \overline Y_1 - \widehat\bbeta_2(\widehat \bmc_2)^\top  \frac{n_1}{n} (\overline{\bmX}_2^{\mathrm {imp}} (\widehat \bmc_2) - \overline{\bmX}_1^{\mathrm {imp}} (\widehat \bmc_1))   + \widehat\bbeta_1(\widehat \bmc_1)^\top  \frac{n_2}{n} (\overline{\bmX}_1^{\mathrm {imp}} (\widehat \bmc_1) - \overline{\bmX}_2^{\mathrm {imp}} (\widehat \bmc_2))  \\
	&=  \overline Y_2  -\overline Y_1 - \left\{\widehat\bbeta_2(\widehat \bmc_2) \frac{n_1}{n}  + \widehat\bbeta_1(\widehat \bmc_1) \frac{n_2}{n}   \right\}^\top (\overline{\bmX}_2^{\mathrm {imp}} (\widehat \bmc_2) - \overline{\bmX}_1^{\mathrm {imp}} (\widehat \bmc_1)).
\end{align*}

As $\widehat \bmc_t  =\bmc+o_p(1)$, 
we have $ \widehat\bbeta_t (\widehat \bmc_t 
) =   \bbeta_t (\bmc ) +o_p(1), $ for $t=1,2$ from Lemma \ref{lemma: beta}.  Note that 
 \begin{align*}
	& \overline{\bmX}_2^{\mathrm {imp}} (\widehat \bmc_2) - \overline{\bmX}_1^{\mathrm {imp}} (\widehat \bmc_1) - \{ 	 \overline{\bmX}_2^{\mathrm {imp}} ( \bmc) - \overline{\bmX}_1^{\mathrm {imp}} ( \bmc) \}\\
	 &= \frac{1}{n_2} \sum_{ i : A_i =a_2} (\bm 1_J- \bmR_i ) \circ (\widehat \bmc_2 - \bmc)    -  \frac{1}{n_1} \sum_{ i : A_i =a_1} (\bm 1_J- \bmR_i ) \circ (\widehat \bmc_1 - \bmc)  \\
	 &=\bigg\{  \frac{1}{n_2} \sum_{ i : A_i =a_2} (\bm 1_J- \bmR_i )  - E( \bm 1_J- \bmR ) \bigg\} \circ (\widehat \bmc_2 - \bmc) \\
	 & \quad   -   \bigg\{ \frac{1}{n_1} \sum_{ i : A_i =a_1} (\bm 1_J- \bmR_i ) - E( \bm 1_J- \bmR ) \bigg\}  \circ (\widehat \bmc_1 - \bmc) +  E( \bm 1_J- \bmR )  \circ (\widehat \bmc_2 -\widehat \bmc_1) \\
 &=   o_p(n^{-1/2}) +E( \bm 1_J- \bmR )  \circ (\widehat \bmc_2 -\widehat \bmc_1) ,
\end{align*}
{where $\circ$ denotes the Hadamard production.}
Hence, we can write 
\begin{align*}
    {\widehat\theta_{{}_{\mathrm{ANHC,2}}}^{\mathrm {ti}}(\widehat \bmc_1, \widehat \bmc_2)-\widehat\theta_{{}_{\mathrm{ANHC,1}}}^{\mathrm {ti}}(\widehat \bmc_1, \widehat \bmc_2)} &=  \overline Y_2  -\overline Y_1 - \left\{\widehat\bbeta_2(\widehat \bmc_2) \frac{n_1}{n}  + \widehat\bbeta_1(\widehat \bmc_1) \frac{n_2}{n}   \right\}^\top (\overline{\bmX}_2^{\mathrm {imp}} (\bmc) - \overline{\bmX}_1^{\mathrm {imp}} (\bmc)) \\
    & \quad - \left\{\widehat\bbeta_2(\widehat \bmc_2) \frac{n_1}{n}  + \widehat\bbeta_1(\widehat \bmc_1) \frac{n_2}{n}   \right\}^\top   E( 1_J- \bmR )  \circ (\widehat \bmc_2 - \widehat \bmc_1) + o_p(n^{-1/2}).
\end{align*}

Now, if $\widehat \bmc_1$ and $\widehat\bmc_2$ are the treatment-specific observed covariate means, i.e., $\widehat \bmc_t=(\widehat c_{t1},\dots, \widehat c_{tJ})^\top $ with $\widehat c_{tj}= \sum_{i: A_i = a_t} R_{ij}X_{ij} /  \sum_{i: A_i = a_t} R_{ij}$ for $t=1,2$, then $\widehat \bmc_2-\widehat \bmc_1=O_p(n^{-1/2} )$. Also because $\overline{\bmX}_2^{\mathrm {imp}} (\bmc) - \overline{\bmX}_1^{\mathrm {imp}} (\bmc) = O_p(n^{-1/2} )$, we can further simplify the expression as 
\begin{align*}
       &{\widehat\theta_{{}_{\mathrm{ANHC,2}}}^{\mathrm {ti}}(\widehat \bmc_1, \widehat \bmc_2)-\widehat\theta_{{}_{\mathrm{ANHC,1}}}^{\mathrm {ti}}(\widehat \bmc_1, \widehat \bmc_2)}\\
        &=  \overline Y_2  -\overline Y_1 - (\pi_1\bbeta_2(\bmc)+\pi_2\bbeta_1(\bmc) )^\top (\overline{\bmX}_2^{\mathrm {imp}} (\bmc) - \overline{\bmX}_1^{\mathrm {imp}} (\bmc)) \\
    & \quad - (\pi_1\bbeta_2(\bmc)+\pi_2\bbeta_1(\bmc) )^\top   E( \bm 1_J- \bmR )  \circ (\widehat \bmc_2 - \widehat \bmc_1) + o_p(n^{-1/2})\\
    &=  \underbrace{\overline Y_2  -\overline Y_1 - (\pi_1\bbeta_2(\bmc)+\pi_2\bbeta_1(\bmc) )^\top (\overline{\bmX}_2^{\mathrm {imp}} (\bmc) - \overline{\bmX}_1^{\mathrm {imp}} (\bmc))}_{M_1} \\
    & \quad - \underbrace{(\pi_1\bbeta_2(\bmc)+\pi_2\bbeta_1(\bmc) )^\top   E( \bm 1_J- \bmR ) \oslash E(\bmR) \circ (\overline{\bmX}_2^{\mathrm {imp}} (\bmc) - \overline{\bmX}_1^{\mathrm {imp}} (\bmc) )}_{M_2} + o_p(n^{-1/2})
\end{align*}
where $\oslash$ denotes Hadamard division, and the last equality is because 
\begin{align*}
    & \overline{\bmX}_2^{\mathrm {imp}} (\bmc) - \overline{\bmX}_1^{\mathrm {imp}} (\bmc) \\
    &= \frac{\sum_{i: A_i =a_2} \bmR_i\circ \bmX_i+(\bm 1_J -\bmR_i) \circ \bmc }{n_2} - \frac{\sum_{i: A_i =a_1} \bmR_i\circ \bmX_i+(\bm 1_J -\bmR_i) \circ \bmc }{n_1} \\
    &= \begin{pmatrix}
    \cdots \\
    \frac{\sum_{i: A_i=a_2} R_{ij} X_{ij} + (1- R_{ij}) c_j }{n_2} -\frac{\sum_{i: A_i=a_1} R_{ij} X_{ij} + (1- R_{ij}) c_j }{n_1}  \\
    \cdots 
    \end{pmatrix}\\
    & =\begin{pmatrix}
    \cdots \\
    (\widehat c_{2j} - c_{j}) \frac{\sum_{i: A_i =a_2} R_{ij} }{n_2} -(\widehat c_{1j} - c_{j}) \frac{\sum_{i: A_i =a_1} R_{ij} }{n_1} \\
    \cdots 
    \end{pmatrix}\\
    &=(\widehat\bmc_2- \bmc) \circ \frac{\sum_{i: A_i =a_2} \bmR_i }{n_2} -(\widehat\bmc_1-\bmc)\circ  \frac{\sum_{i: A_i =a_1} \bmR_i }{n_1} \\
    &= (\widehat\bmc_2- \bmc) \circ  E(\bmR_i) -(\widehat\bmc_1-\bmc)\circ  E(\bmR_i) +o_p(n^{-1/2} ) \\
    &= (\widehat\bmc_2- \widehat\bmc_1) \circ  E(\bmR_i) +o_p(n^{-1/2} ).
\end{align*} 

From Theorem 1 of \cite{ye2020principles}, $M_1$ has the smallest asymptotic variance among all linearly adjusted estimators  and thus has smaller variance compared to $M_1+M_2$. This concludes the proof.

\subsection{Proof of Lemma \ref{lemma: mim as cwi}}

Based on Lemma 3 of \cite{zhao2022adjust}, we know MIM is invariant to the imputation value, so we have $\widehat{\theta}_{{}_{{\mathrm {ANHC}},t}}^{\mathrm {mim}}(\bm 0)=\widehat{\theta}_{{}_{{\mathrm {ANHC}},t}}^{\mathrm {mim}}(\widehat{\bmc}_t^\star)$.

Next, we prove that $\widehat{\theta}_{{}_{{\mathrm {ANHC}},t}}^{\mathrm {mim}}(\bm 0)=\widehat{\theta}_{{}_{{\mathrm {ANHC}},t}}^{\mathrm {cwi}}(\widehat{\bmc}_t^\star)$.  Note that the design matrix for $Y\sim 1+ \impx (\bm 0) + \bmR  $ using data with $A_i=a_t$ is 
\begin{align*}
    \bm D_1 = \begin{pmatrix}
    \bm{1}_{n_t} & (\bmR_{1} \bmX_{1})_{n_t} &  \dots & (\bmR_{J} \bmX_{J})_{n_t}  & (\bmR_{1})_{n_t} & \dots & (\bmR_{J})_{n_t}
    \end{pmatrix}_{n_t\times (2J+1)} ,
\end{align*}
where $(\bm{X}_j \bm{R}_{j})_{n_t}$ is a $n_t$-dimensional column vector of $R_{ij}X_{ij}$'s with $A_i=a_t$, and similarly for $(\bmR_{j})_{n_t}$, $j=1,\dots, J$. Denote the resulting coefficient estimators as $ \widetilde{\bm \tau}_1=( \widetilde{\alpha}_1,  \widetilde{\bbeta}_t(\bm 0)^\top , \widetilde{\bgamma}_t(\bm 0)^\top)^\top$ where $ \widetilde{\bbeta}_t(\bm 0) = ( \widetilde \beta_{t1}(0),\dots,  \widetilde\beta_{tJ}(0))^\top $ and $ \widetilde{\bgamma}_t(\bm 0)=( \widetilde \gamma_{t1}( 0),\dots,  \widetilde\gamma_{tJ}( 0))^\top$ are both $J$-dimensional column vectors. Now consider the design matrix for $Y\sim 1+ \impx ( \widehat \bmc^\star_t) + \bmR  $ using data with $A_i=a_t$, where $ \widehat\bmc_t^\star=- \widetilde{\bgamma}_t(\bm 0)\oslash \widetilde{\bbeta}_t(\bm 0)$:
\begin{align*}
    \bm{D}_2 = \begin{pmatrix}
    \bm{1}_{n_t} & (\bmR_{1} \bmX_{1})_{n_t} + (\bm 1_{n_t}- (\bmR_{1})_{n_t}) \cdot \left( -  \frac{  \widetilde \gamma_{t1} (0)}{ \widetilde \beta_{t1} (0)}\right)  &  \dots  & (\bmR_{1})_{n_t} & \dots 
    \end{pmatrix}
\end{align*}
Denote the resulting coefficient estimators as $ \widetilde{\bm\tau}_2$. Some calculation reveals that we can write $\bm{D}_2= \bm{D}_1 \bm\Gamma$, with 
\begin{align*}
  & \bm\Gamma=\begin{bmatrix}
\bm 1_J & -( \widetilde{\bgamma}_t(\bm 0)\oslash \widetilde{\bbeta}_t(\bm 0))^\top & \bm 0_J^\top\\
\bm 0_J & \bm{I}_{J\times J} & \bm 0_{J\times J}\\
\bm 0_J & {\mathrm {diag}}( \widetilde{\bgamma}_t(\bm 0)\oslash \widetilde{\bbeta}_t(\bm 0)) & \bm I_{J\times J}
\end{bmatrix}_{(2J+1)\times(2J+1)}, 
\end{align*}
where ${\mathrm {diag}}( \widetilde{\bgamma}_t(\bm 0)\oslash \widetilde{\bbeta}_t(\bm 0))$ is a $J\times J$ diagonal matrix whose diagonal elements are those in the vector $ \widetilde{\bgamma}_t(\bm 0)\oslash \widetilde{\bbeta}_t(\bm 0)$,  $\bm I_{J\times J}$ being a $J\times J$ identity matrix, $\bm 0_{J\times J}$ being a $J\times J$ matrix of zeros, and $\bm 0_J$ is a $J$-dimensional vector of zeros. Note that $\bm \Gamma$ is full rank and invertible. By Lemma \ref{lemmas:ding}, we know $ \widetilde{ \bm\tau}_1=\bm \Gamma \widetilde{\bm \tau}_2$. It is straightforward to solve for  $ \widetilde{ \bm\tau}_1=\bm\Gamma \widetilde{\bm\tau}_2$ and obtain that $\widetilde{\bm\tau}_2= ( \widetilde{\alpha}_1+\sum_{j=1}^J \widetilde\gamma_{tj}(0), \widetilde\bbeta_t(\bm 0)^\top , \bm 0^\top )^\top $. By the fact that linear projection is unique, with data from $A_i=a_t$, 
we have that the linear projection of $Y$ onto the space spanned by   $\sim 1+ \impx (\widehat \bmc^\star_t) + \bmR  $ is the same as the projection onto the space spanned by $\sim 1+ \impx (\widehat \bmc^\star_t)  $. This implies that with imputation values $\widehat\bmc^\star_t$, ANHECOVA under CWI is the same as ANHECOVA under MIM, both give the fitted model $\widehat{\mu}_t(\bmX;\widehat{\bmc}_t^\star)= \widetilde{\alpha}_1+\sum_{j=1}^J \widetilde\gamma_{tj}(0)  + \widetilde \bbeta_t (\bm 0)^\top  \impx ( \widehat \bmc^\star_t)    $. This implies that $\widehat{\theta}_{{}_{{\mathrm {ANHC}},t}}^{\mathrm {cwi}}(\widehat{\bmc}_t^\star)=\widehat\theta_{{}_{{\mathrm {ANHC}},t}}^{\mathrm {mim}} (\widehat \bmc_t^\star)$, as desired.

\subsection{Proof of Theorem \ref{theo:asymp}}

First, we have
\begin{align*}
    \widehat{\theta}_t(\widehat \bb_t; \widehat{\bmc}_t)&=\overline{Y}_t-\widehat{\bb}_t^{\top}(\overline{\bmX}_t^{\mathrm{ imp}}(\widehat{\bmc}_t)-\overline{\bmX}^{\mathrm{ imp}}(\widehat{\bmc}_t))\\
    &=\overline{Y}_t-\bb_t^{\top}(\overline{\bmX}_t^{\mathrm{ imp}}(\widehat{\bmc}_t)-\overline{\bmX}^{\mathrm{ imp}}(\widehat{\bmc}_t))+(\bb_t-\widehat{\bb}_t)(\overline{\bmX}_t^{\mathrm{ imp}}(\widehat{\bmc}_t)-\overline{\bmX}^{\mathrm{ imp}}(\widehat{\bmc}_t))\\
    &=\overline{Y}_t-\bb_t^{\top}(\overline{\bmX}_t^{\mathrm{ imp}}(\widehat{\bmc}_t)-\overline{\bmX}^{\mathrm{ imp}}(\widehat{\bmc}_t))+o_p(1/\sqrt{n})\\
    &= \overline{Y}_t-\bb_t^{\top}(\overline{\bmX}_t^{\mathrm{ imp}}({\bmc}_t)-\overline{\bmX}^{\mathrm{ imp}}({\bmc}_t))\\
    &\quad -\bb_t^{\top}((\overline{\bmX}_t^{\mathrm{ imp}}(\widehat{\bmc}_t)-\overline{\bmX}^{\mathrm{ imp}}(\widehat{\bmc}_t))-(\overline{\bmX}_t^{\mathrm{ imp}}({\bmc}_t)-\overline{\bmX}^{\mathrm{ imp}}({\bmc}_t)) ) +o_p(1/\sqrt{n}) ,
\end{align*}
where the third line is because $\overline{\bmX}_t^{\mathrm{ imp}}(\widehat{\bmc}_t)-\overline{\bmX}^{\mathrm{ imp}}(\widehat{\bmc}_t)= O_p(n^{-1/2})$ and $\bb_t-\widehat{\bb}_t=o_p(1)$.

Next,  from 
\begin{align*}
    &(\overline{\bmX}_t^{\mathrm{ imp}}(\widehat{\bmc}_t)-\overline{\bmX}^{\mathrm{ imp}}(\widehat{\bmc}_t)) -(\overline{\bmX}_t^{\mathrm{ imp}}({\bmc}_t)-\overline{\bmX}^{\mathrm{ imp}}({\bmc}_t)) \\
    &= \frac{1}{n_t}\sum_{i:A_i=a_t}(\bmR_i\circ \bmX_i+(\bm 1_{J}-\bmR_i)\circ \widehat{\bmc}_t)-\frac{1}{n}\sum_{i=1}^n(\bmR_i\circ \bmX_i+(\bm 1_{J}-\bmR_i)\circ {\bmc}_t) \\
    & \quad -\frac{1}{n_t}\sum_{i:A_t=a_t}(\bmR_i\circ \bmX_i+(\bm 1_J-\bmR_i)\circ\widehat{\bmc}_t)+\frac{1}{n}\sum_{i=1}^n(\bmR_i\circ \bmX_i+(\bm 1_J-\bmR_i)\circ\bmc_t)\\
    &= \frac{1}{n_t}\sum_{i:A_i=a_t}(\bm 1_J-\bmR_i)\circ (\widehat{\bmc}_t-\bmc_t)-\frac{1}{n}\sum_{i=1}^n (\bm 1_J-\bmR_i)\circ (\widehat{\bmc}_t-\bmc_t)\\
    &= \left(\frac{1}{n_t}\sum_{i:A_t=a_t}(\bm 1_J-\bmR_i)-\frac{1}{n}\sum_{i=1}^n(\bm 1_J-\bmR_i)\right)\circ (\widehat{\bmc}_t-\bmc_t)\\
    &= O_p(1/\sqrt{n})o_p(1)\\
    &=o_p(1/\sqrt{n}) .
\end{align*}
Therefore, we have $    \widehat{\theta}_t(\widehat \bb_t; \widehat{\bmc}_t) =    \widehat{\theta}_t( \bb_t; {\bmc}_t) +o_p(1/\sqrt{n})$. {Write the sample average as $ \mathbb{E}_n [\mu(\bmX) ]= n^{-1} \sum_{i=1}^{n} \mu(\bmX_i) $.}
Let $\bm\mu_X(\bmc_t)= E(\bmX^{\mathrm{ imp}} (\bmc_t))$. Now, we calculate
\begin{align*}
    &\widehat{\bmtheta}(\bb_1, \dots, \bb_k, \bmc_1, \dots, \bmc_k)-\bmtheta  = \begin{pmatrix}
    \overline{Y}_1 - \theta_1 -\bb_1^{\top}(\overline{\bmX}_1^{\mathrm{ imp}}({\bmc}_1) -\overline{\bmX}^{\mathrm{ imp}}(\bmc_1))\\
    \cdots\\
    \overline{Y}_k - \theta_k -\bb_k^{\top}(\overline{\bmX}^{\mathrm{ imp}}_k(\bmc_k)-\overline{\bmX}^{\mathrm{ imp}}(\bmc_k))\\
    \end{pmatrix} \\
    & = \begin{pmatrix}
    \overline{Y}_1 - \theta_1 -\bb_1^{\top}(\overline{\bmX}_1^{\mathrm{ imp}}(\bmc_1)-\bm\mu_X(\bmc_1))\\
    \cdots\\
    \overline{Y}_k - \theta_k -\bb_k^{\top}(\overline{\bmX}_k^{\mathrm{ imp}}(\bmc_k)-\bm\mu_X(\bmc_k))\\
    \end{pmatrix} + \begin{pmatrix}
  \bb_1^{\top}(\overline{\bmX}^{\mathrm{ imp}}(\bmc_1) - \bm\mu_X(\bmc_1))\\
    \cdots\\
     \bb_k^{\top}(\overline{\bmX}^{\mathrm{ imp}}(\bmc_k) - \bm\mu_X(\bmc_k))\\
    \end{pmatrix}\\
    &= \underbrace{\begin{pmatrix}
    \pi_1^{-1}\mathbb{E}_n\left[I(\bmA=\bma_1)\left(Y-\theta_1-\bb_1^{\top}(\bmX^{\mathrm{ imp}}(\bmc_1)-\bm\mu_X(\bmc_1))\right)\right]\\
    \cdots\\
    \pi_k^{-1}\mathbb{E}_n\left[I(\bmA=\bma_k\left(Y-\theta_k-\bb_k^{\top}(\bmX^{\mathrm{ imp}}(\bmc_k)-\bm\mu_X(\bmc_k))\right)\right]
    \end{pmatrix}}_{\bm M_1} + \underbrace{\begin{pmatrix}
    \bb_1^{\top} \mathbb{E}_n[\bmX^{\mathrm{ imp}}(\bmc_1)-\bm\mu_X(\bmc_1)]\\
    \cdots\\
    \bb_k^{\top} \mathbb{E}_n[\bmX^{\mathrm{ imp}}(\bmc_k)-\bm\mu_X(\bmc_k)]
    \end{pmatrix}}_{\bm M_2} \\
    & \quad + {o_p(n^{-1/2} )} .
\end{align*}
Asymptotic normality follows from a similar argument as in the proof of Theorem 1 in \cite{ye2020principles}. It remains to calculate the asymptotic variance. 

Note that $n \var (\bm{M}_1)= {\mathrm {diag}} \{  \pi_t^{-1} \var (Y^{(t)} - \bb_t ^\top \impx (\bmc_1) ) \} $. For $\bm{M}_2$, we reformulate it as 
 \begin{displaymath}
  \bm{M}_2= \underbrace{\left(
    \begin{array}{ccc}
      \bb_1^{\top}&& \\
           &\ddots&\\
           && \bb_k^{\top}
    \end{array}\right)}_{B}
 \begin{pmatrix}
    \mathbb{E}_n[\bmX^{\mathrm{ imp}}(\bmc_1)-\bm\mu_X(\bmc_1)]\\
    \cdots\\
    \mathbb{E}_n[\bmX^{\mathrm{ imp}}(\bmc_k)-\bm\mu_X(\bmc_k)]
    \end{pmatrix} .
\end{displaymath}
Hence, $ n \var (\bm{M}_2)= \bm{B}\bm\Sigma_X(\bmc_1,\dots,\bmc_k)\bm{B}^\top $, where {\small $$\bm\Sigma_X(\bmc_1,\dots,\bmc_k)=\begin{pmatrix}
\var(\bmX^{\mathrm{ imp}}(\bmc_1)) & \cov(\bmX^{\mathrm{ imp}}(\bmc_1),X^{\mathrm{ imp}}(\bmc_2)) & \cdots & \cov(\bmX^{\mathrm{ imp}}(\bmc_1),\bmX^{\mathrm{ imp}}(\bmc_k))\\
\cov(\bmX^{\mathrm{ imp}}(\bmc_2),\bmX^{\mathrm{ imp}}(\bmc_1))& \var(\bmX^{\mathrm{ imp}}(\bmc_2)) & \cdots & \cov(\bmX^{\mathrm{ imp}}(\bmc_2),\bmX^{\mathrm{ imp}}(\bmc_k))\\
\vdots & \vdots & \ddots & \vdots\\
\cov(\bmX^{\mathrm{ imp}}(\bmc_k),\bmX^{\mathrm{ imp}}(\bmc_1)) & \cov(\bmX^{\mathrm{ imp}}(\bmc_k),\bmX^{\mathrm{ imp}}(\bmc_2))& \cdots & \var(\bmX^{\mathrm{ imp}}(\bmc_k))
\end{pmatrix}$$}

The $(t,s)$ element in $\cov (\bm M_1, \bm M_2) $ is 
{\small \begin{align*}
    &\cov\bigg(\pi_t^{-1}\mathbb{E}_n\left[I(\bmA=\bma_t)\left(Y-\theta_t-(\bmX^{\mathrm{ imp}}(\bmc_t)-\bm\mu_X(\bmc_t))^{\top}\bb_t\right)\right], \bb_s^{\top}\sum_{t=1}^k\mathbb{E}_n\left[I(\bmA=\bma_t)(\bmX^{\mathrm{ imp}}(\bmc_s)-\bm\mu_X(\bmc_s))\right]\bigg)\\
    &= \cov\bigg(\pi_t^{-1}\mathbb{E}_n\left[I(\bmA=\bma_t)\left(Y-\theta_t-(\bmX^{\mathrm{imp}}(\bmc_t)-\bm\mu_X(\bmc_t))^{\top}\bb_t\right)\right], \bb_s^{\top}\mathbb{E}_n\left[I(\bmA=\bma_t)(\bmX^{\mathrm{ imp}}(\bmc_s)-\bm\mu_X(\bmc_s))\right]\bigg)\\
    &= n^{-1}\pi_t^{-1}\cov\left(I(\bmA=\bma_t)(Y-\bmX^{\mathrm{ imp}}(\bmc_t)^{\top}\bb_t), \bb_s^{\top}I(\bmA=\bma_t)(\bmX^{\mathrm{ imp}}(\bmc_s)-\bm\mu_X(\bmc_s))\right)\\
    &= n^{-1}\pi_t^{-1}E\left(I(\bmA=\bma_t)(Y-\bmX^{\mathrm{ imp}}(\bmc_t)^{\top}\bb_t)\bb_s^{\top}(\bmX^{\mathrm{ imp}}(\bmc_s)-\bm\mu_X(\bmc_s))\right)\\
    &=n^{-1}E\left(\left(Y^{(t)}-\bmX^{\mathrm{ imp}}(\bmc_t)^{\top}\bb_t\right)\bb_s^{\top}\left(\bmX^{\mathrm{ imp}}(\bmc_s)-\bm\mu_X(\bmc_s)\right)\right)\\
    &=n^{-1}\left(\cov\left(Y^{(t)}, \bb_s^{\top}\bmX^{\mathrm{ imp}}(\bmc_s)\right)-\cov\left(\bmX^{\mathrm{ imp}}(\bmc_t)^{\top}\bb_t,\bb_s^{\top}\bmX^{\mathrm{ imp}}(\bmc_s)\right)\right)\\
    &=n^{-1}\left\{\bbeta_{t}(\bmc_s)^{\top} \var(\bmX^{\mathrm{ imp}}(\bmc_s)) \bb_{s}-\bb_{t}^{\top} \cov(\bmX^{\mathrm{ imp}}(\bmc_t),\bmX^{\mathrm{ imp}}(\bmc_s)) \bb_{s}
    \right\} .
\end{align*}}
Thus, we have $n \cov(\bm{M}_1, \bm{M}_2)=\mathscr{B} \bm{D}_X (\bmc_1,\dots,  \bmc_k) \bm{B}^\top -\bm{B}\bm\Sigma_X(\bmc_1,\dots,\bmc_k)B^\top$. Similarly, 
$n \cov (\bm{M}_2, \bm{M}_1)=\bm{B}\bm{D}_X (\bmc_1,\dots,  \bmc_k)\mathscr{B}^\top -\bm{B}\bm\Sigma_X(\bmc_1,\dots,\bmc_k)\bm{B}^\top $.

Summing up all the terms gives the desirable results.

\subsection{Proof of Corollary \ref{theo: asymp.cwi}.}

Note that when estimating linear contrast of response means using CWI, our estimator is $( \widehat\theta^{\mathrm{cwi}}_{_{\mathrm{ANHC},t}} (\widehat \bmc_t) -   \widehat\theta^{\mathrm{cwi}}_{_{\mathrm{ANHC},s}} (\widehat \bmc_s)) - (\theta_t-\theta_s) = (\bma_t - \bma_s)^\top\{ \widehat{\theta}(\bb_1, \dots, \bb_k, \bmc_1, \dots, \bmc_k)-\theta  \}  $ {where $\bma_t$ is a vector with the $t$th entry being 1 while the rest being 0 and similarly for $\bma_s$}. By Theorem \ref{theo:asymp} and delta method, we have 
\begin{align}
		&\sqrt{n} \big(       \widehat\theta^{\mathrm{cwi}}_{_{\mathrm{ANHC},t}} (\widehat \bmc_t) -   \widehat\theta^{\mathrm{cwi}}_{_{\mathrm{ANHC},s}} (\widehat \bmc_s)  - (\theta_t - \theta_s) \big)  \xrightarrow{d} N\big( 0, \sigma_{{{ts}}}^2(\bmc_t,\bmc_s) \big),  \label{eq: anhc}
	\end{align}
with 
\begin{align*}
\sigma_{{{ts}}}^2(\bmc_t,\bmc_s) &= (\bma_t - \bma_s)^\top V(\bm{B}; \bmc_1,\dots, \bmc_k) (\bma_t - \bma_s) \\
&=  (\bma_t - \bma_s)^\top   {\mathrm {diag}} \{  \pi_t^{-1} \var (Y^{(t)} - \bb_t ^\top \impx (\bmc_t) ) \}    (\bma_t  - \bma_s) \\
&\quad- (\bma_t - \bma_s)^\top \bm{B}\bm\Sigma_X(\bmc_1,\dots,\bmc_k)\bm B^\top (\bma_t - \bma_s) \\
&\quad + (\bma_t - \bma_s)^\top \mathscr{B}\bm D_X(\bmc_1,\dots,\bmc_k)\bm{B}^\top(\bma_t - \bma_s)\\
&\quad+(\bma_t - \bma_s)^\top \bm{B}\bm D_X(\bmc_1,\dots,\bmc_k)\mathscr{\bmB}^\top (\bma_t - \bma_s)\\
&=\pi_t^{-1} \var (Y^{(t)} - \bb_t ^\top \impx (\bmc_t) ) +\pi_s^{-1} \var (Y^{(s)} - \bb_s ^\top \impx (\bmc_s) ) \\
&\quad - \bb_t^\top\var(\impx(\bmc_t))\bb_t - \bb_s^\top\var(\impx(\bmc_s))\bb_s  +2 \bb_t^\top \cov(\impx(\bmc_t), \impx(\bmc_s))\bb_s\\
&\quad +\bmbeta_t(\bmc_t)^\top \var(\impx(\bmc_t))\bb_t+\bmbeta_s(\bmc_s)^\top\var(\impx(\bmc_s))\bb_s-2\bbeta_s(\bmc_t)^\top \var(\impx(\bmc_t))\bb_t \\
&\quad - 2\bmbeta_t(\bmc_s)^\top \var(\impx(\bmc_s))\bb_s +\bb_t^\top\var(\impx(\bmc_t))\bmbeta_t(\bmc_t)+\bb_s^\top\var(\impx(\bmc_s))\bmbeta_t(\bmc_s)
\end{align*}

Note that $\bb_t=\bbeta_t(\bmc_t)$. Hence the above expression simplifies to 
\begin{align*}
\sigma_{{{ts}}}^2(\bmc_t,\bmc_s) &=\pi_t^{-1} \var (Y^{(t)} - \bmbeta_t(\bmc_t) ^\top \impx (\bmc_t) ) +\pi_s^{-1} \var (Y^{(s)} - \bmbeta_s (\bmc_s)^\top \impx (\bmc_s) ) \\
&\quad +2 \bmbeta_t(\bmc_t)^\top \cov(\impx(\bmc_t), \impx(\bmc_s))\bmbeta_s(\bmc_s)+\bmbeta_t(\bmc_t)^\top \var(\impx(\bmc_t))\bmbeta_t(\bmc_t)\\
&\quad +\bmbeta_s(\bmc_s)^\top\var(\impx(\bmc_s))\bmbeta_s(\bmc_s)-2\bbeta_s(\bmc_t)^\top \var(\impx(\bmc_t))\bmbeta_t(\bmc_t) \\
&\quad - 2\bmbeta_t(\bmc_s)^\top \var(\impx(\bmc_s))\bmbeta_s (\bmc_s)
\end{align*}


\subsection{Proof of Theorem \ref{theorem:optimality}}

First we show that the MIM used in combination with the CWI is invariant to the CWI values $\widehat \bmc_1, \dots, \widehat \bmc_k$. We apply Lemma 3 in \cite{zhao2022adjust} to each
$\widehat\theta_{{}_{{\mathrm{ANHC}},t}}^{{\mathrm{mim}}} (\widehat\bmc_t) =   \overline Y_t  - \widetilde \bbeta_{t}(\widehat\bmc_t)^\top \{ \overline{\bmX}_{t}^{\mathrm{imp}} (\bm 0) -  \overline{\bmX}^{\mathrm{imp}} (\bm 0)\} - \widetilde {\bgamma}_{t}(\widehat\bmc_t)^\top ( \overline{\bmR}_t -  \overline{\bmR}) $ and obtain that $\widehat\theta_{{}_{{\mathrm{ANHC}},t}}^{{\mathrm{mim}}} (\widehat\bmc_t) $ is invariant to the imputation value $\widehat \bmc_t $, so that $\widehat\theta_{{}_{{\mathrm{ANHC}},t}}^{{\mathrm{mim}}} (\widehat\bmc_t)= \widehat\theta_{{}_{{\mathrm{ANHC}},t}}^{{\mathrm{mim}}} (\bm 0)$. Therefore, the whole vector $$\widehat\theta_{{}_{{\mathrm{ANHC}}}}^{{\mathrm{mim}}} (\widehat\bmc_1,\dots, \widehat \bmc_k) = (\widehat\theta_{{}_{{\mathrm{ANHC}},1}}^{{\mathrm{mim}}} (\widehat\bmc_1), \dots, \widehat\theta_{{}_{{\mathrm{ANHC}},k}}^{{\mathrm{mim}}} (\widehat\bmc_k))^\top = (\widehat\theta_{{}_{{\mathrm{ANHC}},1}}^{{\mathrm{mim}}} (\bm 0), \dots, \widehat\theta_{{}_{{\mathrm{ANHC}},k}}^{{\mathrm{mim}}} (\bm 0))^\top $$ and is invariant to imputation vector $\widehat{\bmc}_1,\dots,\widehat{\bmc}_k$. This implies that even in the CWI framework, there is only one ANHECOVA MIM estimator and thus it suffices to consider $\widehat\theta_{{}_{{\mathrm{ANHC}}}}^{{\mathrm{mim}}} (\bm 0)$.  This result extends Lemma 3 in \cite{zhao2022adjust} to the CWI framework. 

Next, we show that the class of estimators $\mathcal{S}_1= \{ \widehat{\bmtheta}_{\mathrm L}({\bb}_1, \dots,  \bb_k, {\br}_1,\dots, {\br}_k; {\bmc}_1, \dots, {\bmc}_k): \text{for any } {\bb}_t, \br_t, {\bmc}_t\}$ is the same as the class $\mathcal{S}_2= \{ \widehat{\bmtheta}_{\mathrm L}({\bb}_1, \dots,  \bb_k, {\br}_1,\dots, {\br}_k; \bm 0, \dots, \bm 0): \text{for any } {\bb}_t, \br_t\}$. It is obvious that $\mathcal{S}_2\subset \mathcal{S}_1$. It remains to show that the reverse is also true. Note that for any vector ${\bm\tau}_t^\top=(\bb_t^\top,{\bm r}_t^\top)$, we have 
$$\widehat{\theta}_t(\bb_t, \bm r_t, \bmc_t)=\begin{pmatrix}
\overline{Y}_t-\bm\tau_t^{\top}
\begin{pmatrix}\overline{\bmX}^{\mathrm{imp}}_t(\bmc_t)-\bm\mu_X(\bmc_t)\\\overline{\bmR}_t-\bm\mu_R\end{pmatrix}
\end{pmatrix}+\begin{pmatrix}
{\bm\tau}_t^{\top}\begin{pmatrix}\overline{\bmX}^{\mathrm{imp}}(\bmc_t)-\bm\mu_X(\bmc_t)\\\overline{\bmR}-\bm\mu_R\end{pmatrix}
\end{pmatrix}. 
$$
Since 
\begin{align*}
    \overline{\bmX}^{\mathrm{imp}}_t(\bmc_t)&=\frac{\sum_{i:A_i=a_t} (\bmX_i\circ \bmR_i+\bmc_t\circ(\bm 1_J-\bmR_i))}{n_t}\\
    &=\frac{\sum_{i:A_i=a_t} \bmX_i\circ \bmR_i}{n_t}+\frac{\sum_{i:A_i=a_t} (\bm 1_J-\bmR_i)\circ \bmc_t}{n_t}\\
    &=\overline{\bmX}^{\mathrm{imp}}_t(\bm 0)+\bmc_t\circ\overline{\bm M}_t
\end{align*}
where $\overline{ \bm M}_t=\bm 1_J-\overline{\bmR}_t$. Let $\bm\mu_M = \bm 1_J - \bm\mu_R $, we have $\overline{ \bm M}_t-\bm\mu_M=-(\overline{\bmR}_t-\bm\mu_R)$. Thus, we have 
\begin{align*}
    \begin{pmatrix}\overline{\bmX}^{\mathrm{imp}}_t(\bmc_t)-\bm\mu_X(\bmc_t)\\\overline{\bmR}_t-\bm\mu_R\end{pmatrix} &= 
    \begin{pmatrix}
    	\overline{\bmX}^{\mathrm{imp}}_t(\bm 0)+\bmc_t\circ \overline{ \bm M}_t-\bm\mu_X(\bmc_t)\\\overline{\bmR}_t-\bm\mu_R
    \end{pmatrix}\\
    &=\begin{pmatrix}\overline{\bmX}^{\mathrm{imp}}_t(\bm0)+\bmc_t\circ\overline{\bm M}_t-\bm\mu_X(\bm 0)-\bmc_t\circ\bm\mu_M\\\overline{\bmR}_t-\bm\mu_R\end{pmatrix}\\
    &=\begin{pmatrix}\overline{\bmX}^{\mathrm{imp}}_t(\bm 0)-\bm\mu_X(\bm 0)+\bmc_t\circ(\overline{\bm M}_t-\bm\mu_M)\\\overline{\bmR}_t-\bm\mu_R\end{pmatrix}\\
    &=\begin{pmatrix}\overline{\bmX}^{\mathrm{imp}}_t(\bm 0)-\bm\mu_X(\bm 0)-\bmc_t\circ(\overline{\bmR}_t-\bm\mu_R)\\\overline{\bmR}_t-\bm\mu_R\end{pmatrix}_{2J\times 1}\\
    &=\begin{pmatrix}
    \bm I_{J\times J} & \mathrm{diag} (-\bmc_t) \\
    \bm 0_{J\times J} & \bm I_{J\times J} 
    \end{pmatrix}
\begin{pmatrix}
    \overline{\bmX}^{\mathrm{imp}}_t(\bm 0)-\bm\mu_X(0)\\\overline{\bmR}_t-\bm\mu_R
    \end{pmatrix} .
\end{align*}

Similarly, we have $\overline{\bmX}^{\mathrm{imp}}(\bmc_t)=\overline{\bmX}^{\mathrm{imp}}(\bm 0)+\bmc_t\circ\overline{\bm M}$ and 
$$\begin{pmatrix}\overline{\bmX}^{\mathrm{imp}}(\bmc_t)-\bm\mu_X(\bmc_t)\\\overline{\bmR}-\bm\mu_R\end{pmatrix}=\begin{pmatrix}
	\bm I_{J\times J} & \mathrm{diag} (-\bmc_t) \\
	\bm 0_{J\times J} & \bm I_{J\times J} 
\end{pmatrix}
\begin{pmatrix}
    \overline{\bmX}^{\mathrm{imp}}(\bm 0)-\bm\mu_X(\bm 0)\\\overline{\bmR}-\bm\mu_R
    \end{pmatrix}. 
$$
Hence, 
\begin{align*}
	\widehat{\theta}_t(\bb_t, \bm r_t, \bmc_t) & =\begin{pmatrix}
		\overline{Y}_t-\bm \tau_t^{\top}
	\begin{pmatrix}
		\bm I_{J\times J} & \mathrm{diag} (-\bmc_t) \\
		\bm 0_{J\times J} & \bm I_{J\times J} 
	\end{pmatrix}
	\begin{pmatrix}
		\overline{\bmX}^{\mathrm{imp}}_t(\bm 0)-\bm\mu_X(\bm 0)\\\overline{\bmR}_t-\bm\mu_R
	\end{pmatrix} 
	\end{pmatrix}\\
	&\quad + \begin{pmatrix}
		{\bm\tau}_t^{\top}
		\begin{pmatrix}
			\bm I_{J\times J} & \mathrm{diag} (-\bmc_t) \\
			\bm 0_{J\times J} & \bm I_{J\times J} 
		\end{pmatrix}
		\begin{pmatrix}
			\overline{\bmX}^{\mathrm{imp}}(\bm 0)-\bm \mu_X(\bm 0)\\\overline{\bmR}-\bm \mu_R
		\end{pmatrix} 
	\end{pmatrix}\\
&= \begin{pmatrix}
	\overline{Y}_t-\check{\bm{\tau}}_t^{\top}
	\begin{pmatrix}
		\overline{\bmX}^{\mathrm{imp}}_t(\bm 0)-\bm \mu_X(\bm 0)\\\overline{\bmR}_t-\bm \mu_R
	\end{pmatrix} 
\end{pmatrix} + 
\begin{pmatrix}
\check{\bm\tau}_t^{\top}
	\begin{pmatrix}
		\overline{\bmX}^{\mathrm{imp}}(\bm 0)-\bm\mu_X(\bm 0)\\
		\overline{\bmR}-\bm\mu_R
	\end{pmatrix} 
\end{pmatrix} ,
\end{align*}
where $\check{\bm \tau}^\top = \bm\tau_t^\top 		\begin{pmatrix}
	\bm I_{J\times J} & \mathrm{diag} (-\bmc_t) \\
	\bm 0_{J\times J} & \bm I_{J\times J} 
\end{pmatrix}$. Therefore, we see that any estimator 
$\widehat{\theta}_t(\bb_t, \bm r_t, \bmc_t)$ in $\mathcal{S}_1$  belongs to the set $\mathcal{S}_2$. In conclusion, $\mathcal{S}_1= \mathcal{S}_2$. Theorem 1 of \cite{ye2020principles} implies that $\widehat\bmtheta_{{}_{{\mathrm{ANHC}}}}^{{\mathrm{mim}}} (\bm 0)$ achieves optimal efficiency within the $\mathcal{S}_2$, and thus $\widehat\bmtheta_{{}_{{\mathrm{ANHC}}}}^{{\mathrm{mim}}} (\bm 0)$  is also optimal in $\mathcal{S}_1$.  Because we showed in Lemma  \ref{lemma: mim as cwi} that $\anhc^{{\mathrm {mim}}} (\bm 0) $ actually belongs to the smaller class $\{ \widehat{\bmtheta}({\bb}_1, \dots,  \bb_k; {\bmc}_1, \dots, {\bmc}_k): \text{for any } {\bb}_t, {\bmc}_t\}$, it must be optimal within this small class, which concludes the proof.

\subsection{Proof of Theorem \ref{lemma:equi}}

First, we present the explicit form of $\lim_{n\rightarrow \infty}\widetilde{\gamma}_{tj}(0)$ and $\lim_{n\rightarrow \infty}\widetilde{\beta}_{tj}(0)$. We will omit the $\mathrm{imp}$ superscript  below to simplify the notations. From 
$$
\lim_{n\rightarrow\infty}\begin{pmatrix}
	\widetilde{\bbeta}_t (\bm 0)\\ \widetilde{\bm\gamma}_t (\bm 0) \end{pmatrix}=
\var\left(\begin{pmatrix}\bmX (\bm 0)\\ \bmR
\end{pmatrix}\right)^{-1}\cov\left(\begin{pmatrix}\bmX(\bm 0)\\ \bmR\end{pmatrix}, Y\mid \bmA=\bma_t\right), 
$$ and using the formula of inverse of block matrices, we can get
{\small \begin{align*}
		& \lim_{n\rightarrow\infty}\widetilde{\bbeta}_{t} (\bm 0) = \left(\var(\bmX(\bm 0))-\cov(\bmX(\bm 0),\bmR^\top)\var(\bmR)^{-1}\cov(\bmR,\bmX(\bm 0)^\top)\right)^{-1} \cov(\bmX(\bm 0),Y^{(t)})\\
		&\quad-\var(\bmX(\bm 0))^{-1}\cov(\bmX(\bm 0),\bmR^\top) \left( \var(\bmR)-\cov(\bmR, \bmX(\bm 0)^\top)\var(\bmX(\bm 0))^{-1}\cov(\bmX(\bm 0), \bmR^{\top}) \right)^{-1}\cov(\bmR,Y^{(t)})\\
		& \lim_{n\rightarrow\infty} \widetilde{\bm\gamma}_{t} (\bm 0)=-\var(\bmR)^{-1}\cov(\bmR,\bmX(\bm 0)^\top)\left(\var(\bmX(\bm 0))-\cov(\bmX(\bm 0),\bmR^\top)\var(\bmR)^{-1}\cov(\bmR,\bmX(\bm 0)^\top)\right)^{-1}\\
		&\quad \cov(\bmX(\bm 0),Y^{(t)}) + \left( \var(\bmR)-\cov(\bmR, \bmX(\bm 0)^\top)\var(\bmX(\bm 0))^{-1}\cov(\bmX(\bm 0),  \bmR^{\top}) \right)^{-1}\cov(\bmR,Y^{(t)}) .
\end{align*}}
Now, we prove the theorem. By the assumption that $\lim_{n\to\infty}\widetilde{\gamma}_{tj}(0)/\lim_{n\to\infty}\widetilde{\beta}_{tj}(0)$ does not depend on $t$ for every $j$, we have that the probability limit of $\widehat{\bmc}_t^\star$ does not depend on $t$. We denote the probability limit as ${\bmc}^\star$. By Lemma \ref{lemma: mim as cwi}, we have $\widehat{\bmtheta}_{{}_{ \mathrm{ANHC},t}}^{ \mathrm{mim}} (\bm 0) =\widehat{\bmtheta}_{{}_{ \mathrm{ANHC},t}}^{ \mathrm{cwi}}(\widehat\bmc_t^\star)$. By Theorem \ref{theo:asymp}, we have 
\[
\widehat{\bmtheta}_{{}_{ \mathrm{ANHC},t}}^{ \mathrm{mim}} (\bm 0) = 
\widehat{\bmtheta}_{{}_{ \mathrm{ANHC},t}}^{ \mathrm{cwi}}(\widehat\bmc_t^\star) = \widehat{\bmtheta}_{{}_{ \mathrm{ANHC},t}}^{ \mathrm{cwi}}(\bmc^\star) +  o_p(1/\sqrt{n})
\]
Note that since $\bmc^\star$ does not depend on $t$, single imputation using any $\widehat \bmc$ such that $\widehat \bmc \xrightarrow{p} \bmc^\star$ leads to an asymptotically equivalent estimator, i.e., $\widehat{\bmtheta}_{{}_{ \mathrm{ANHC},t}}^{ \mathrm{si}}(\widehat \bmc)  =   \widehat{\bmtheta}_{{}_{ \mathrm{ANHC},t}}^{ \mathrm{cwi}}(\bmc^\star) +  o_p(1/\sqrt{n})$.

{\color{black}{\subsection{Proof of Corollary \ref{corollary:MCAR}}
When the data is missing completely at random, we have $\bmR\perp (\bmX, \bmA, Y^{(t)})$ for all $t$. Notice that when {$\bmR\perp Y^{(t)}$}, the formula of $\lim_{n\rightarrow\infty}\widetilde{\bbeta}_{t} (\bm 0) $ and $\lim_{n\rightarrow\infty} \widetilde{\bm\gamma}_{t} (\bm 0)$ in Section S2.7 can be simplified to 
{\begin{align*}
		\lim_{n\rightarrow\infty}\widetilde{\bbeta}_{t} (\bm 0)&= \left(\var(\bmX(\bm 0))-\cov(\bmX(\bm 0),\bmR^\top)\var(\bmR)^{-1}\cov(\bmR,\bmX(\bm 0)^\top)\right)^{-1}\cov(\bmX(\bm 0),Y^{(t)})\\
		\lim_{n\rightarrow\infty}\widetilde{\bm \gamma}_{t} (\bm 0)&=-\var(\bmR)^{-1}\cov(\bmR,\bmX(\bm 0)^\top)\left(\var(\bmX(\bm 0))-\cov(\bmX(\bm 0),\bmR^\top)\var(\bmR)^{-1}\cov(\bmR,\bmX(\bm 0)^\top)\right)^{-1}\\
		&\quad \cov(\bmX(\bm 0),Y^{(t)}) =-\var(\bmR)^{-1}\cov(\bmR,\bmX(\bm 0)^\top)\lim_{n\rightarrow\infty}\widetilde{\bbeta}_{t} (\bm 0).
\end{align*}}

First, we will show $\cov(\bmR,\bmX(\bm 0)^\top)=\var(\bmR)\text{diag}(E(\bmX_j))$. We know $\cov(\bmR,\bmX(\bm 0)^\top)=\cov(\bmR, (\bmX\circ\bmR)^\top )$, and since $\bmR\perp\bmX$
\begin{align*}
(\cov(\bmR, (\bmX\circ\bmR)^\top ))_{ij} &=\cov(\bmR_i, \bmX_j\bmR_j) \\
&= E(\bmX_j\bmR_i\bmR_j)-E(\bmX_j\bmR_j)E(\bmR_i)\\
&= E(\bmX_j)\cov(\bmR_i, \bmR_j)\\
&= E(\bmX_j)\var(\bmR)_{ij}
\end{align*}
so $\cov(\bmR,\bmX(\bm 0)^\top)=\cov(\bmR, (\bmX\circ\bmR)^\top )=\var(\bmR)\text{diag}(E(\bmX_j))$.

Then, we have $$-\var(\bmR)^{-1}\cov(\bmR,\bmX(\bm 0)^\top)= -\var(\bmR)^{-1}\var(\bmR)\text{diag}(E(\bmX_j))=-\text{diag}(E(\bmX_j))$$ where $\text{diag}(E(\bmX_j))$ is a $J\times J$ diagonal matrix with the $j$-th diagonal value $E(\bmX_j)$.
Therefore, we have $\widehat{\bmc}_t^*=-\lim_{n\rightarrow\infty}\widetilde{\bm \gamma}_t (\bm 0) \oslash\lim_{n\rightarrow\infty}\widetilde{\bbeta}_t (\bm 0)=(\var(\bmR)^{-1}\cov(\bmR,\bmX(\bm 0)^\top)\lim_{n\rightarrow\infty}\widetilde{\bbeta}_{t} (\bm 0))\oslash \lim_{n\rightarrow\infty}\widetilde{\bbeta}_t (\bm 0) =\text{diag}(E(\bmX_j))\lim_{n\rightarrow\infty}\widetilde{\bbeta}_{t} (\bm 0)\oslash \lim_{n\rightarrow\infty}\widetilde{\bbeta}_t (\bm 0) = E(\bmX)=E(\bmX\mid \bmR=1)$}}.

\subsection{Asymptotic limit of $  - \lim_{n\to \infty}\widetilde \gamma_{t} (0)/ \lim_{n\to \infty}\widetilde \beta_{t} (0)
	$ with a single covariate}



When $ \var\begin{pmatrix} XR\\R\end{pmatrix}$ is invertible, 
\begin{align*}
  \lim_{n\rightarrow\infty}  \begin{pmatrix} \widetilde \beta_{t} (0) \\ \widetilde \gamma_{t} (0)\end{pmatrix} &=\left( \var\begin{pmatrix} XR\\R\end{pmatrix}\right)^{-1}\cov\left(\begin{pmatrix} XR\\R\end{pmatrix}, Y\mid A=a_t\right)\\
    &= \begin{pmatrix}
    \var(XR) &\cov(XR, R)\\
   \cov(XR, R) & \var(R)
    \end{pmatrix}^{-1}\begin{pmatrix}
   \cov(XR, Y\mid A=a_t)\\\cov(R, Y\mid A=a_t)
    \end{pmatrix}\\
    &= \frac{1}{\Delta}\begin{pmatrix}
    \var(R) & -\cov(XR,R)\\
    -\cov(XR,R) & \var(XR)
    \end{pmatrix}\begin{pmatrix}
   \cov(XR, Y\mid A=a_t)\\\cov(R, Y\mid A=a_t)
    \end{pmatrix}\\
    &= \frac{1}{\Delta}\begin{pmatrix}
    \var(R)\cov(XR, Y\mid A=a_t)-\cov(XR,R)\cov(R, Y\mid A=a_t)\\
    \var(XR)\cov(R, Y\mid A=a_t)-\cov(XR,R)\cov(XR, Y\mid A=a_t)
    \end{pmatrix} ,
\end{align*}
where $\Delta=\var(XR)\var(R)-\cov(XR,R)^2$. Hence, 
\begin{align*}
	- \frac{\lim_{n\rightarrow\infty} \widetilde \gamma_t (0)}{\lim_{n\rightarrow\infty} \widetilde \beta _t (0)} =  - \frac{\var (XR) \cov (R, Y^{(t)}) - \cov (XR, R) \cov (XR, Y^{(t)})}{\var (R) \cov (XR, Y^{(t)}) - \cov (XR, R) \cov (R, Y^{(t)})} . 
\end{align*}

\section{Optimal SI value with a single covariate when $R\not\perp (Y^{(1)},Y^{(2)})$} \label{supp:opt.nind}
When $R$ is dependent on $(Y^{(1)}, Y^{(2)})$, we can find the optimal $c$ by minimizing $\sigma^2_{_{\mathrm{ANHC}}}(c)$ or equivalently by maximizing 
$$
\sigma^2_{_{\mathrm{AN}}}-\sigma^2_{_{\mathrm{ANHC}}}(c)= \frac{ \{\cov (RX, \pi_1Y^{(2)}+ \pi_2Y^{(1)} ) +\cov (1-R, \pi_1Y^{(2)}+ \pi_2Y^{(1)} ) c \}^2  }{\pi_1 \pi_2 \var (RX + (1-R) c) }.
$$

Taking the derivative of our objective function $M(c)$ with respect to $c$, we note that the numerator of the derivative can be written as $(a_1c+b_1)(a_2c+b_2)$, while the denominator of the derivative is positive, where $a_1=2\pi_1\pi_2\tau_R$,  $a_2=-\tau_{RX}\cdot \var(R)-\tau_R\cdot \var(R)\E(X|R=1)$, $b_1=2\pi_1\pi_2\tau_{RX}$, and $b_2=\tau_R\cdot \{\E(R)\var(X\mid R=1)+\var(R)\E(X\mid R=1)^2\}+\tau_{RX}\cdot \var(R)\E(X|R=1)$,  $\tau_{RX}=\cov(RX,\pi_1Y^{(2)}+\pi_2Y^{(1)})$ and $\tau_R=\cov(1-R,\pi_1Y^{(2)}+\pi_2Y^{(1)})$.  Letting the derivative equal 0, we get the two roots 
\begin{align*}
    c_1 & =-\frac{\tau_{RX}}{\tau_R}, \\
    c_2 & =\frac{\tau_R\cdot \{\E(R)\var(X|R=1)+\var(R)\E(X|R=1)^2\}+\tau_{RX}\cdot \var(R)\E(X|R=1)}{\var(R)\E(X|R=1)\cdot \tau_R+\var(R)\cdot \tau_{RX}}.
\end{align*}

\begin{figure}[h]
    \centering
    \includegraphics[width=\linewidth]{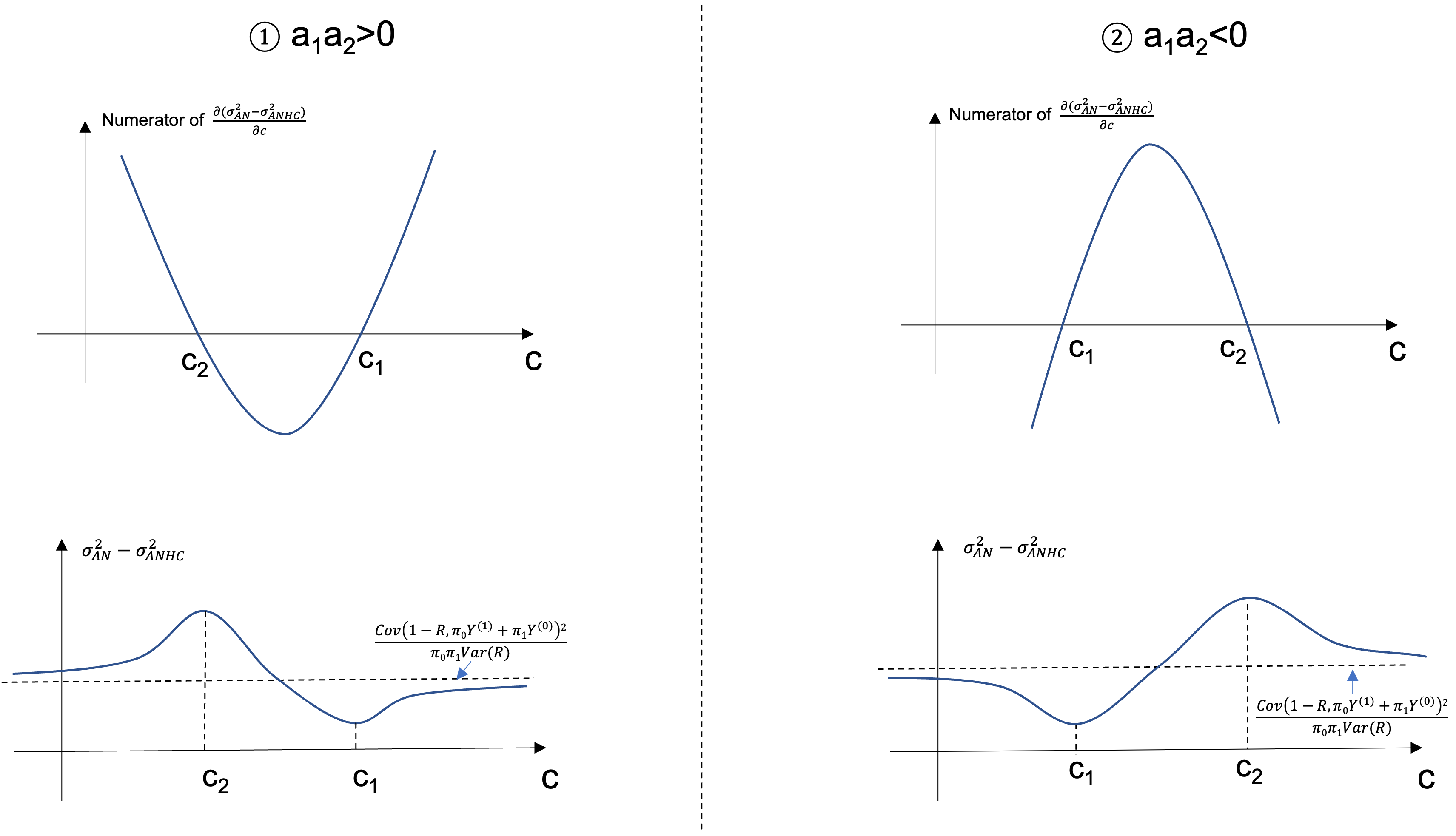}
    \caption{Illustration of the relationship between $c_1$ and $c_2$ when $a_1a_2>0$ or $a_1a_2<0$}
    \label{fig:univar}
\end{figure}

Now, we will show $c_2$ is the global maximum of $\sigma^2_{_{\mathrm{AN}}}-\sigma^2_{_{\mathrm{ANHC}}}(c)$. Since the denominator of the derivative is positive, it suffices to show the numerator of the derivative, $N=(a_1c+b_1)(a_2c+b_2)$, satisfies (as in Figure \ref{fig:univar})
\begin{enumerate}
    \item When $a_1a_2>0$, $c_2<c_1$; When $a_1a_2<0$, $c_2>c_1$.
    \item $\ans^2-\anhcs^2 (c)$ has the same limit as $c\to \infty$ and $c\to -\infty$, so it can't be as large as what $c_2$ reaches.
\end{enumerate}

Using the expression of $c_1$ and $c_2$, we have 
\begin{align*}
    c_2-c_1 &= \frac{\tau_R\cdot\{\E(R)\var{(X|R=1)}+\var(R)\E(X|R=1)^2\}+\tau_{RX}\cdot\var{(R)}\E(X|R=1)}{\var(R)\E(X|R=1)\cdot \tau_R+\var{(R)}\cdot \tau_{RX}}+\frac{\tau_{RX}}{\tau_R}\\
    & = \frac{\tau_{RX}\cdot \{\tau_R\cdot \var(R)\E(X|R=1)+\tau_{RX}\cdot \var(R)\}+\tau_R^2\cdot\{\E(R)\var(X|R=1)\}}{\tau_R\{\tau_R\cdot\var(R)\E(X|R=1)+\tau_{RX}\cdot \var(R)\}}\\
    & \quad + \frac{\var(R)\E(X|R=1)^2+\tau_{RX}\tau_R\cdot\E(X|R=1)\var(R)}{\tau_R\{\tau_R\cdot\var(R)\E(X|R=1)+\tau_{RX}\cdot \var(R)\}} .
\end{align*}
Looking at the numerator of $c_2-c_1$, we have 
\begin{align*}
    &\tau_{RX}\{\tau_R\var(R)\E(X|R=1)+\tau_{RX} \var(R)\}+\tau_R^2\{\E(R)\var(X|R=1)+\var(R)\E(X|R=1)^2\}\\
    &+\tau_{RX}\tau_R\ \E(X|R=1)\var(R)\\
    &=\tau_{RX}\cdot\var{(R)}\{\tau_R\cdot\E(X|R=1)+\tau_{RX}\}+\tau_R\cdot \var(R)\E(X|R=1)\{\tau_{R}\cdot\E(X|R=1)+\tau_{RX}\}\\
    &+\tau_R^2\  \E(R)\var(X|R=1)\\
    &=\var(R)\{\tau_{RX}+\tau_R\cdot \E(X|R=1)\}^2+\tau_R^2\  \E(R)\var(X|R=1)>0
\end{align*}

Notice that \begin{align*}
	a_1a_2 & =-2\pi_1\pi_2\tau_R\{\tau_{RX}\cdot\var(R)+\tau_{R}\cdot\var(R)\E(X|R=1)\}\\
	&=-2\pi_1\pi_2\times\{\text{denominator of } (c_2-c_1) \} 
\end{align*}

We have obtained that when $a_1a_2>0$, we have $c_2<c_1$, and when $a_1a_2<0$, we have $c_2>c_1$. As $\ans^2-\anhcs^2 (c_1) = 0$, we know that $c_1$ is the global minima of the objective function; as shown in Figure \ref{fig:univar}, we know that $c_2$ is a local maxima  of the objective function. In fact, $c_2$ is a global maxima of the objective function because the objective function achieves the same limit as $c\to \infty$ and $c\to -\infty$ by applying L'Hospital's rule, 
\begin{align*}
 &   \lim_{|c|\rightarrow \infty} \frac{ \{\cov (RX, \pi_1Y^{(2)}+ \pi_2Y^{(1)} ) +\cov (1-R, \pi_1Y^{(2)}+ \pi_2Y^{(1)} ) c \}^2  }{\pi_1\pi_2 \var (RX + (1-R) c) }\\
    &=\frac{\cov(1-R,\pi_1Y^{(2)}+\pi_2Y^{(1)})^2}{\pi_1\pi_2\var{(R)}}.
\end{align*}

We want to note that when $\tau_{RX}=-\tau_R\cdot \E(X|R=1)$, we have $a_1a_2=0$. In this case, {$a_1 b_2 + b_1 a_2= 2\pi_1\pi_2\tau_R^2 E(R)\var(X\mid R=1)>0$}, $c_2$ is not well defined, and $c_1$ becomes $E(X\mid R=1)$. {Plugging in $\tau_{RX}=-\tau_R\cdot \E(X|R=1)$ to the objective function, we get 
\begin{align*}
M(c)&=\frac{\tau_R^2(E(X\mid R=1)-c)^2}{\pi_1\pi_2E(R)\var(X\mid R=1)+\pi_1\pi_2(E(X\mid R=1)-c)^2\var(R)},
\end{align*}
which is obvious that $c_1=E(X\mid R=1)$ is the minimum of this objective function and $M(c)=0$. }This means that in this particular scenario, the common practice of imputing using $E(X\mid R=1)$ is the least efficient choice.

In summary, with a single covariate, when $\tau_{XR}+\tau_R\cdot \E(X|R=1)\neq 0$, imputing using $c_2$ leads to the optimal efficiency; when $\tau_{XR}+\tau_R\cdot \E(X|R=1)= 0$, imputing using a pre-specified large number leads to near optimal efficiency. The results above can be generalized to the case when we have $J>1$ but $(R_1, X_1) \perp \cdots \perp (R_J, X_J)$.